\documentclass[pre,twocolumn,showpacs,amsmath,amssymb]{revtex4}
\usepackage{graphicx}
\usepackage{dcolumn}
\usepackage{bm}

\begin{document}

\title{Environmental effects in the quantum-classical transition for the delta-kicked harmonic oscillator}
\author{A.R.R. Carvalho$^1$, R. L. de Matos Filho$^2$, and L. Davidovich$^2$}
\affiliation{$^1$Max-Planck-Institut f\"ur Physik komplexer Systeme,
N\"othnitzer Strasse 38, D-01187 Dresden \\$^2$Instituto de F\'{\i}sica,
Universidade Federal do Rio de Janeiro, Caixa Postal 68528, 21941-972 Rio de Janeiro, RJ, Brazil}

\date{\today}

\begin{abstract}
We discuss the roles of the macroscopic limit and of different system-environment
interactions in the quantum-classical transition for a chaotic system. We
consider the kicked harmonic oscillator subject to reservoirs that correspond in the classical case to purely dissipative or purely diffusive behavior, in a situation that can be implemented in ion trap experiments. In
the dissipative case, we derive an expression for the time at which quantum
and classical predictions become different (breaking time) and show that
a complete quantum-classical correspondence is not possible in the chaotic
regime. For the diffusive environment we estimate the minimum value of the 
diffusion coefficient necessary to retrieve the classical limit and also
show numerical evidence that, for diffusion below this threshold, the breaking 
time behaves, essentially, as in the case of the system without a reservoir.
\end{abstract}

\pacs{05.45.Mt, 03.65.Yz, 05.45.-a}
\maketitle

\section{INTRODUCTION}

The problem of understanding the classical world from quantum theory is subtle and especially challenging when dealing with
classically chaotic systems.

Even the definition of classical chaos cannot be directly translated to
quantum mechanics. Indeed, exponential sensitivity to initial conditions, used
to define classical chaos, relies on the concept of individual trajectories
in phase space, which is absent in quantum formalism. The use of classical
phase space distributions, instead of trajectories, seems to be the way to
circumvent this problem, since they can be readily compared with
quasi-probability distributions defined for the corresponding quantum system. 

One expects, however, that the dynamics of the quantum and the corresponding
classical system should differ, after some time, even if the initial
distributions coincide. This time, often called Ehrenfest
time or breaking time, while large for integrable systems, can be very short for chaotic systems. Indeed, in this case it has been shown~\cite{log} to be proportional to the
logarithm of the inverse of an effective Planck constant, $\hbar_{\rm eff}$,
which is the ratio between Planck constant and a typical action of the
system. For integrable systems, on the other hand, it scales as an inverse power of $\hbar_{\rm eff}$. In fact, quantum corrections become important when the distribution is
able to explore the non-linearities of the potential, which can occur in a
logarithmic time scale due to the exponentially fast stretching of the
distribution, imposed by chaotic dynamics. 

One way to face this problem is to go to the macroscopic limit, namely
$\hbar_{eff} \rightarrow 0$, resulting in an infinite breaking 
time $\tau_{\hbar}$. Nevertheless, for any physical system, $\hbar_{eff}$ is not zero and
therefore $\tau_{\hbar}$ has a finite value, which can be short, even for macroscopic
systems. Indeed, it has been argued that, due to the shortness of the separation time, even components of the solar system would exhibit quantum features, which is in contradiction with observation~\cite{Hyperion,Hyperion3}.

Reconciliation of quantum and classical predictions in this case is provided by the
irreversible coupling of the system with an environment~\cite{feynman,caldeira1,caldeira2}, which leads to the
elimination of the quantum signatures, so that quantum and classical
evolutions remain alike. In systems that, isolated, exhibit dynamical localization, it was
shown~\cite{ott,dittrich,wilkinson,cohen} that noise and dissipation can strongly alter
the situation and under certain conditions restore the classical-like momentum
diffusion. For particular choices of environment, it has been also shown that this reconciliation is possible under some conditions that involve a
scaling relation between the effective Planck constant, the non-linearity
parameter and the strength of the system-reservoir
interaction~\cite{ott,kolovsky,cohen,zurek,pattquantum,pattparameter,habib}. 

One of the aims of this paper is to further explore the conjoint role of
the macroscopic limit ($\hbar_{eff}\ll1$) and of the interaction with the 
environment in the quantum-classical correspondence. In particular, we are
also interested in the regions where the conditions for classicality do
not hold and examine the time scales at which the quantum-classical
correspondence breaks down. We should emphasize that this time scale is
different from the one present in the localization problem studied
in~\cite{ott,dittrich,wilkinson,cohen} since this phenomenom can be absent in the model
we are considering~\cite{rebuzzini}. Another goal of this
paper is to analyze in detail the impact of different system-environment
couplings within the framework of a model that can be implemented experimentally. 

For this purpose, we revisit the kicked harmonic oscillator (KHO), which has
 been the subject of studies both in the classical~\cite{weakchaos} and the quantum
 description~\cite{nonlinear4,rebuzzini,bambi}. Despite some peculiarities and numerical
 difficulties presented by the KHO, the possibilities of implementation with
current available technology in ion traps~\cite{gardiner} turn this model into a
very attractive one. Moreover, in ion traps, one is able to create
artificial reservoirs~\cite{poyatos} and, in particular, different kinds of
 system-environment interactions have already been produced
 experimentally~\cite{turchette}. This favorable scenario becomes complete
 with the possibility of tuning the effective Planck constant by changing
 experimentally accessible parameters. Indeed, $\hbar_{eff}$ and its scaling
 properties are related to the so-called Lamb-Dicke parameter (to be defined later in this paper), which can be modified
 either by changing the trap frequency or the directions of the laser beams
 interacting with the ion. 

We analyze two limiting cases of environment coupling: zero temperature, which
leads in the classical limit to dissipation without diffusion, and a reservoir that leads to diffusion
without dissipation. This is not an unrealistic situation: indeed, the first
may be mimicked by the sideband cooling mechanism in ion traps, under proper conditions, while the second
corresponds to a white-noise position-independent random force, coupled to the
oscillator. This is known to be the most important source of decoherence in actual
experiments~\cite{turchette}. We present analytical and numerical results
concerning the ``distance'' between quantum and classical predictions and the
breaking time. More specifically, in the dissipative case, expressions for the
breaking time in three different parameter regions are derived and their physical
consequences are discussed. In particular, in the most interesting region, we
have a result similar to the one obtained recently, using another method, by Iomin
and Zaslavsky~\cite{iomin}. For the diffusive environment we establish the minimum value of the diffusive constant in order to restore the classical predictions, and we provide numerical evidence that
the breaking time behaves like the one for the system without reservoir if the diffusion constant is kept below this threshold. We also show that the purely diffusive
reservoir has a much stronger impact on the quantum-classical correspondence than the dissipative one.

In Section \ref{CDyn} we present the main features of the classical model, both in
the absence and the presence of coupling with the environment. In Section \ref{Qdyn}, we introduce the quantum model, discussing 
its connections with the experimental realization in ion traps and with the
corresponding classical model. Section \ref{Results} is divided into three parts, 
showing the results for the system without a reservoir or interacting with
dissipative or diffusive reservoirs. Two appendices contain detailed derivations of some of the results presented in the main text.

\section{CLASSICAL DYNAMICS: THE $\delta$-KICKED HARMONIC OSCILLATOR}
\label{CDyn}
The classical $\delta$-kicked harmonic oscillator has been studied for both
the isolated~\cite{kho_zaslavsky} (without reservoir) and dissipative~\cite{kho_diss} cases. Here we review the basics features of these
models and present also the effects of the interaction with a diffusive environment.

\subsection{System without reservoir}

We consider a particle of mass $m$ in a harmonic potential submitted to a
sequence of periodically applied $\delta$-like pulses. The Hamiltonian
that describes this situation is
\begin{equation}
\label{HAM_cl}
H = \frac{p^2}{2m} + \frac{m \nu^2 x^2}{2} + A \cos(kx) \sum_{n}^{\infty} \delta(t-n\tau),
\end{equation}
where $\nu$ is the oscillator frequency, $\tau$ the interval between two
consecutive kicks and $A$ their amplitude. The kicking potential is position
dependent with a periodicity given by the wave vector $k$.

\begin{figure}
\includegraphics[width=8.5cm]{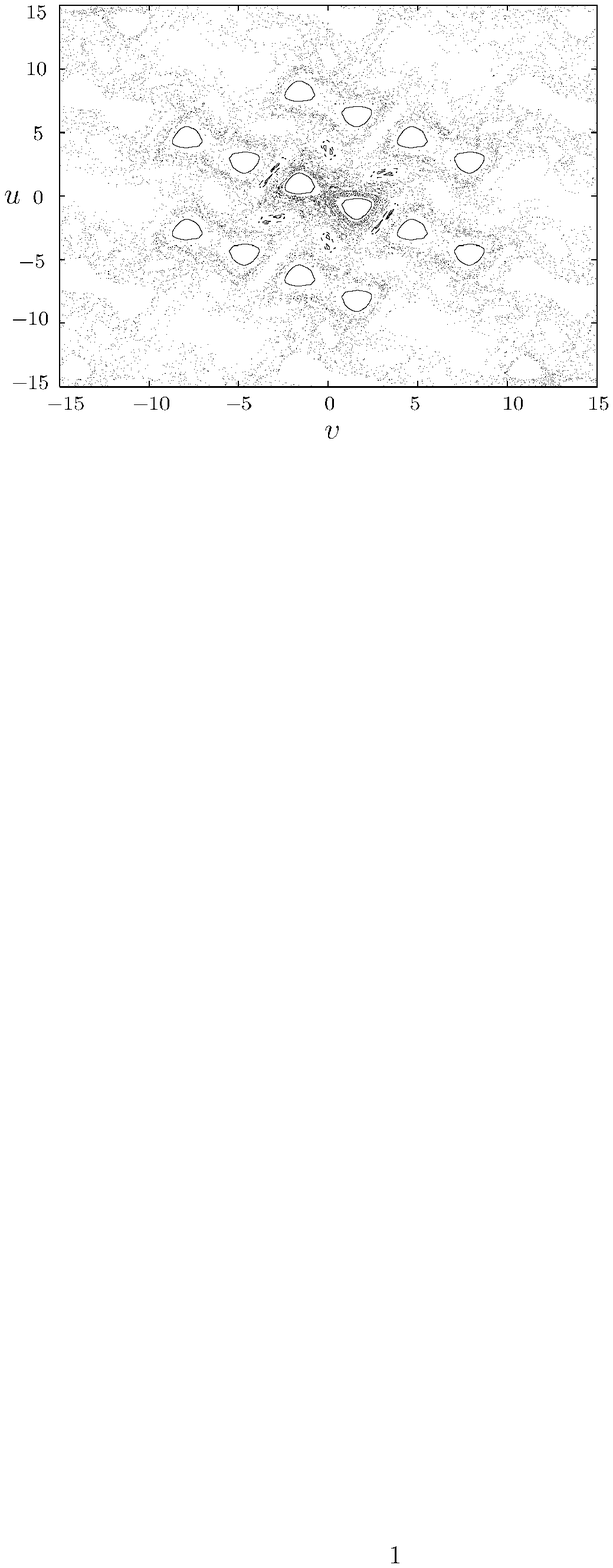}
\caption{Stroboscopic plot for $q=6$ and $K=2.0$, showing some stable
islands as well as the stochastic web forming a hexagonal symmetry in the
unbounded phase space. }
\label{figmapcons}
\end{figure}

The differential equations of motion due to this Hamiltonian can be replaced
by a discrete map. Between two kicks the system evolves accordingly to
\begin{equation}
\label{difeqcons}
\ddot x + \nu^2 x = 0,
\end{equation}
while at the kicking times $n\tau$ there is just a shift in the momentum, so that
\begin{equation}
\label{kickshift}
x_n^+ = x_n, \:\:\:\: p_n^+= p_n +A k \sin(kx),
\end{equation}
where the variables immediately after and before a kick are indicated,
respectively, by the presence or absence of the $+$ superscript. After this integration we can connect the solutions before each kick by the following map 
\begin{subequations}
\begin{equation}
x_{n+1}=\cos(\nu \tau) x_n + \sin(\nu \tau)/m\, \nu \left[p_n +
A\,k\,\sin(k x_n)\right]\,,
\end{equation}
\begin{equation}
p_{n+1}= -m\,\nu\,\sin(\nu \tau) x_n +\cos(\nu\tau)\left[p_n+ A\,k\,\sin(k x_n)\right].
\end{equation}
\end{subequations}
Using dimensionless variables $v$ and $u$ defined by
\begin{eqnarray}
\label{dimvar}
v&=&k x\,, \nonumber\\
u&=&k p/m \nu \, , 
\end{eqnarray}
the map becomes
\begin{subequations}
\label{mapcons}
\begin{equation}
v_{n+1}=\cos(\alpha) v_n + \sin(\alpha) \left[u_n + K \sin(v_n)
\right], 
\end{equation}
\begin{equation}
u_{n+1}=-\sin(\alpha) v_n + \cos(\alpha)\left[u_n + K \sin(v_n) \right],
\end{equation}
\end{subequations}
where $K= A k^2/m \nu$ and $\alpha=\nu\tau$ are, respectively, the
renormalized kicking strength and the ratio between the period of the kicks and the
period of the oscillator. The system's phase space is unbounded and mixed, 
exhibiting stable islands surrounded by a stochastic web along which the
system diffuses. The web is characterized by its thickness that broadens
(shrinks) as the value of the chaoticity parameter
$K$ increases (decreases). For $\alpha=2\pi/q$ ($q$ integer), the stochastic web displays a crystal ($q \in q_c\equiv\{3,\,4,\,6\}$) or quasi-crystal symmetry($q\ne \,q_c$). These basic features
can be seen in Fig.~\ref{figmapcons}, where a stroboscopic plot for the map~(\ref{mapcons}) is shown for
$q=6$ and $K=2.0$ for different initial conditions.

Another useful tool to study the dynamics is to follow the time evolution of the phase-space probability distribution. This method is especially suitable
to problems where the notion of a single deterministic trajectory is absent as
in the case of noisy and quantum dynamics. For the numerical evaluation of the density dynamics, one usually evolves an ensemble of trajectories generated accordingly to the initial distribution and then, by counting the fraction of trajectories lying in each cell of phase space, one recovers the density at a given time. However, this method presents some drawbacks in our case, due to the unboundedness of the phase space and the consequent escape of trajectories. One way to get rid of this problem is to extend the phase space boundaries to be sure that, for the time scale one wants to simulate, no trajectories are lost. Nevertheless, the increase of the phase space area, keeping the size of the cells constant, requires a larger number of trajectories in order to get good statistics. This point imposes severe constraints for an efficient numerical implementation. 

Alternatively, we start with a uniformly distributed ensemble of trajectories,
each of them carrying its own weight related to the initial
distribution. Accordingly to this, the probability at each point is obtained by requiring that 
\begin{equation}
P_n(v_n,u_n)=P_0(v_0,u_0),
\end{equation}
which means that the value of the initial probability $P_0$ at every point
 ($v_0$, $u_0$) in phase space is transported to the image ($v_n$, $u_n$) of
 this point under the action of map~(\ref{mapcons}) after $n$ iterations. 
All the classical quantities calculated throughout the paper are obtained
 through the evaluation of each individual trajectory and then averaging over
 all of them taking into account the respective probabilities. It is important to mention that, although differences between trajectory-based and truly density evolutions are expected~\cite{lasota,fox}, our simulations show no difference between the two methods for sufficiently small phase-space partitions. 

\begin{figure}
\includegraphics[width=8.8cm]{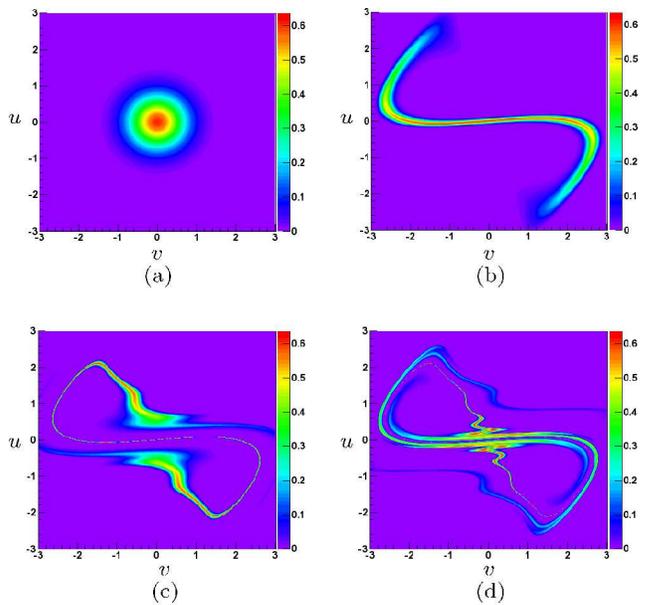}
\caption{(Color online) Classical probability distribution for $q=6$ and $K=2.0$ after 0
(a), 3(b), 6(c) and 9(d) kicks. The same structure of the trajectory-based
stroboscopic map (Fig.~\ref{figmapcons}) is shown. After $9$ kicks only the
central structure is visible. The whole web structure would be seen for larger number of kicks.}
\label{figPDcons}
\end{figure}

Fig.~\ref{figPDcons} displays the evolution of an initial Gaussian
 probability distribution centered at the origin for the same parameters of
 Fig.~\ref{figmapcons}. The numerical procedure to plot the distributions is
 similar to the one described to calculate the averages but, in order not to have problems with the dispersion of trajectories coming from neighbor regions in phase space at $t=0$, we do the calculation backwards in time, choosing the grid at any instant of time $t$ and evolving the points using the inverse map to find the probability of the inverse image of this point at $t=0$.

From Fig.~\ref{figPDcons} one can identify the inner structure already present in Fig.~\ref{figmapcons}. The whole web structure would be visible for larger number of kicks.

\subsection{The dissipative case}

The dynamics of the kicked oscillator changes if a dissipation mechanism is
introduced. In our model this can be achieved by modifying the equation of
motion between the kicks~(\ref{difeqcons}) with the addition of a friction term
proportional to the velocity
\begin{equation}
\label{difeqdiss}
\ddot x + \nu^2 x + \Gamma \dot x = 0,
\end{equation}
where $\Gamma$ is the dissipation rate. 
The discrete map obtained by the integration of~(\ref{difeqdiss}) and the use
of the shift in momentum~(\ref{kickshift}) is 
\begin{subequations}
\begin{equation}
x_{n+1}=e^{-\Gamma\tau/2}\left\{\cos(\bar\alpha) x_n +
\frac{\sin(\bar\alpha)}{m\, \Omega}\left[p_n + Ak\,\sin(k x_n)\right]\right\}, 
\end{equation}
\begin{eqnarray}
p_{n+1}=e^{-\Gamma\tau/2}\Big\{-m\Omega\,\sin(\bar\alpha) x_n+cos(\bar\alpha) \nonumber \\
\times\left[p_n+Ak\,\sin(k x_n)\right]\Big\},
\end{eqnarray}
\end{subequations}
where
\begin{eqnarray}\label{Om}
\Omega&=&\sqrt{\nu^2-\Gamma^2/4}\, ,\nonumber\\
p_n/m&=&\dot x + \Gamma x/2\, , \nonumber \\
\bar\alpha&=&\Omega\tau \, .
\end{eqnarray}

We perform again a change to dimensionless variables
$v'$ and $u'$, so that now
\begin{eqnarray}
\label{dimvar2}
v'&=&k x\,, \nonumber\\
u'&=&k p/m \Omega \, , 
\end{eqnarray}
and
\begin{subequations}
\label{mapaclassicodiss}
\begin{equation}
v'_{n+1} = e^{-\Gamma\tau/2}\left\{\cos(\bar\alpha) v'_n + \sin(\bar\alpha) \left[u'_n + K' \sin(v'_n)
\right]\right\}\,,
\end{equation}
\begin{equation} 
u'_{n+1} = e^{-\Gamma\tau/2}\left\{-\sin(\bar\alpha) v'_n + \cos(\bar\alpha)\left[u'_n + K' \sin(v'_n) \right]\right\},
\end{equation}
\end{subequations}
where $K'=Ak^2/m\Omega$.

\begin{figure}
\includegraphics[width=7.6cm]{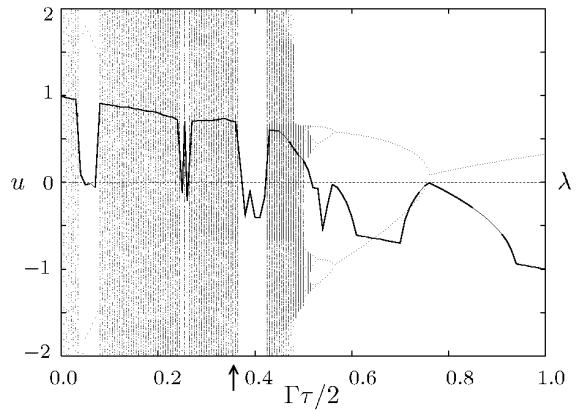}
\caption{Average Lyapunov exponent $\lambda$ (solid line) and bifurcation diagram (dots)
as a function of the dissipation parameter $\Gamma\tau/2$ for $K'=6.0$ and
$q=6$. The horizontal line at $u=0$ was plotted as a reference for the Lyapunov
exponent. For the bifurcation diagram, the vertical axis corresponds to the
the last $10^3$ points of $u$ after $10^6$ iterations of the
map~(\ref{mapaclassicodiss}). We only show the region from $-2$ to $2$ for
better visualization. The arrow refers to the case plotted in Fig.~\ref{figmapdiss}.}
\label{figlyapbif}
\end{figure}

\begin{figure}
\includegraphics[width=7.6cm]{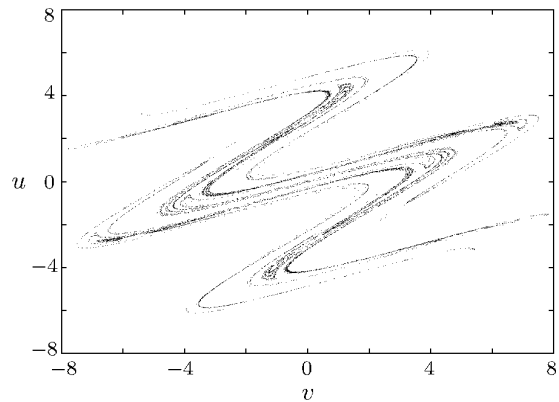}
\caption{Strange attractor obtained from the map~(\ref{mapaclassicodiss}) for
$q=6$, $K'=6.0$ and $\Gamma\tau/2=0.36$ corresponding to a Lyapunov exponent of
$0.697$ in Fig~\ref{figlyapbif} (denoted by an arrow there).}
\label{figmapdiss}
\end{figure}

There are new scenarios arising from the addition of dissipation depending on
the values of $K'$ and $\Gamma\tau$. When one of these parameters is changed, the
system may change from a periodic to a chaotic motion in a sequence of period-doubling bifurcations~\cite{lieberman}. In Fig.~\ref{figlyapbif} we show this sequence and
also the average Lyapunov exponent for $K'=6.0$ and $\Gamma\tau/2$ varying from 0 to
1. For the bifurcation diagram we iterated the map~(\ref{mapaclassicodiss})
for $10^6$ steps and plotted the last $10^3$ points corresponding to the $u$
variable on the vertical axis. The Lyapunov exponent is averaged over $10^4$ different initial
conditions, equally distributed around the origin, taking into account the same
initial probability distribution as in Fig.~\ref{figPDcons}. For each
trajectory the exponent is calculated using the procedure described
in~\cite{lyap} for $10^6$ iterations. One can clearly identify the periodic regions corresponding to non-positive Lyapunov exponents and the chaotic ones where the system goes to strange attractors as the one shown in
Fig.~\ref{figmapdiss}. 

\subsection{The diffusive case}

Another kind of external disturbance that can affect the oscillator dynamics
is the diffusion generated by noise due, for example, to fluctuating forces
acting on the system. One possible mathematical description for the evolution
of the probability distribution between two consecutive kicks is the Fokker-Planck equation 
\begin{equation}
\label{eq:FPcl}
\frac{\partial P}{\partial t} = \nu u \frac{\partial P}{\partial v}
-\nu v \frac{\partial P}{\partial u} + {\cal D} \left( \frac{\partial^2
P}{\partial v^2} + \frac{\partial^2
P}{\partial u^2}\right).
\end{equation}
The first two terms describe the harmonic evolution while the third accounts
for diffusion, both in $v$ and $u$, with diffusion coefficient ${\cal D}$. This diffusive
term has two different effects on the system's dynamics. First, the noise
limits the development of small-scale structures in phase space generated
by the nonlinear dynamics, smoothing out the probability
distribution. While the stretching of the distribution tends to generate thin structures in phase space, noise will tend to counterbalance this effect, leading to a lower limit in the width of such structures, which will depend on both the
nonlinearity parameter and the noise strength \cite{zurek}.
The second effect is a faster spread of the system over phase space. In fact, the diffusion produced by the noise adds a new mechanism
for connecting different parts of the web, thus enhancing the original chaotic
diffusion.

\section{QUANTUM DYNAMICS: quantum $\delta$-kicked harmonic oscillator}
\label{Qdyn}

\subsection{System without reservoir}

The quantum Hamiltonian for the $\delta$-kicked harmonic oscillator is given by replacing in Eq.~(\ref{HAM_cl}) the variables $x$ and $p$ by operators:
\begin{equation}
\hat H = \frac{{\hat p}^2}{2m}+ \frac{m \nu^2 {\hat x}^2}{2} + \hbar K_q
\cos\left(k \hat x \right) \sum_{n}^{\infty} \delta(t-n\tau)\,,
\end{equation}
where we have defined $K_q=A/\hbar$.
 
It was shown in~\cite{gardiner} that this model describes the center-of-mass
motion of an ion in a one-dimensional harmonic trap submitted to a sequence of
standing-wave laser pulses, off-resonance with respect to a
transition between the ground state and another electronic level. In this
off-resonance condition the excited state is negligibly populated and can be
eliminated adiabatically. The result of this elimination is an equation
describing just the motional dynamics subject to recoils associated to the
incoming laser pulses. 

In terms of the annihilation and creation operators $\hat a$
and ${\hat a}^\dag$ for the harmonic oscillator,
\begin{equation}
\label{annihop}
\hat a = \sqrt{{m \nu / 2 \hbar}}\,{\hat x} +i
\sqrt{{1/2 \hbar m \nu}}\, \hat p,
\end{equation} 
the above Hamiltonian can be written as
\begin{equation}
\label{HAM_q}
\hat H = \hbar \nu {\hat a}^\dag \hat a + \hbar K_q \cos\left[\eta (\hat a +
{\hat a}^\dag) \right] \sum_{n}^{\infty} \delta(t-n\tau)\, ,
\end{equation}
where 
\begin{equation}
\label{eta}
\eta = k \sqrt{\hbar/2 m \nu}
\end{equation}
is a scaling parameter related to the macroscopic limit - the so-called Lamb-Dicke parameter~\cite{wineland}. This parameter can be expressed as $\eta=2\pi\Delta x_0/\lambda$, where $\Delta x_0$ is the width of the ground state of the harmonic oscillator, and $\lambda=2\pi/k$ is the wavelength of the kicking force. This is the classicality parameter for the model under consideration. Its square is seen to be the ratio between $\hbar$ and the action $2m\nu/k^2$, playing the role of the dimensionless parameter $\hbar_{eff}$ mentioned in the Introduction. The limit $\eta\rightarrow0$ can be achieved by letting $k\rightarrow0$. This parameter can be easily changed in ion trap experiments, by varying the direction of the kicking laser pulses with respect to the trap axis, since $k$ stands, in this case, for the projection of the lasers wave vectors on the trap axis. 

In terms of $\eta$, we can write:
\begin{eqnarray}
\langle \hat a\rangle &=& \frac{1}{2 \eta} (v + i u)\equiv (\bar v + i \bar u)\,,\\
 K_q&=&\frac{K}{2\eta^2}\,, 
\end{eqnarray}
where~(\ref{dimvar}), (\ref{annihop}) and~(\ref{eta}) were used. While the classicality parameter
$\eta$ appears naturally in the quantum model, when this is expressed in terms of the annihilation and creation operators, it may also be introduced classically, by using the new variables $\bar v$ and $\bar u$, which yields the following scaled map
\begin{subequations}
\label{mapaescalado}
\begin{equation}
\label{mapaescalado1}
\bar v_{n+1}\, = \cos(\alpha) \bar v_n + \sin(\alpha) \left[\bar u_n +
 \frac{K}{2\eta} \sin(2 \eta \bar v_n)
 \right]\,, 
\end{equation}
\begin{equation}
\label{mapaescalado2}
 \bar u_{n+1}\, = - \sin(\alpha) \bar v_n + \cos(\alpha) \left[\bar u_n +
 \frac{K}{2\eta} \sin(2 \eta \bar v_n)\right]. 
\end{equation}
\end{subequations}

On the quantum level, the evolution dictated by~(\ref{HAM_q}) can be written as a map connecting the
state of the system before each consecutive kick as
\begin{eqnarray}
\label{qmap}
\vert \psi \rangle_{n+1} &=& \hat U_{h} \hat U_{k} \vert \psi \rangle_n
\nonumber \\
&=&e^{-i \nu \tau \hat a^\dag \hat a} e^{-i K_q\cos[\eta(\hat a +\hat
a^\dag)]} \vert \psi\rangle_n,
\end{eqnarray}
where $\hat U_{h}$ and $\hat U_{k}$ are, respectively, the evolution operators
for the harmonic oscillator and for the kicks.

\subsection{Open system: influence of environment}

The influence of the environment on the system can be described by the master equation 
\begin{equation}
 \label{eq:lindblad}
 \frac{d\hat{\rho}}{dt}=-\frac{i}{\hbar} \left[\hat H, \hat{\rho} \right] +
 {\cal L}\hat\rho \, ,
\end{equation}
where $\hat{\rho}$ is the reduced density operator of the system in the
interaction picture. The first term on the right hand side
of Eq.~(\ref{eq:lindblad}) corresponds to the unitary dynamics while the second term 
represents the non-unitary effect of the environment in the \emph{Lindblad form} 
\begin{equation}\label{lindbladian}
 {\cal L}\hat\rho \equiv \sum_i(\gamma_i/2)\left(2\,\hat{c}_i\,\hat{\rho}\,\hat{c}_i^\dagger -
\hat{c}_i^\dagger\,\hat{c}_i\,\hat{\rho} -
\hat{\rho}\,\hat{c}_i^\dagger\,\hat{c}_i\right) \, .
\end{equation}
The operators $\hat{c}_i$ are related to the form of
system-environment couplings and the constants $\gamma_i$ measure the strength of these
couplings. 

Equation (\ref{lindbladian}) is frequently found in the description of dissipative systems. It can be derived under very general assumptions, namely Markovicity and complete positivity of the time-evolution of the reduced density operator of system~\cite{lindblad1,lindblad2}. This last condition is defined in the following way. Let $A$ be the system for which the reduced density operator is defined, ${\cal H}_A$ the corresponding Hilbert space, and $\Lambda_A$ the time-evolution map for the reduced density operator $\rho$. Consider any possible extension of ${\cal H}_A$ to the tensor product ${\cal H}_A\otimes{\cal H}_B$, where ${\cal H}_B$ is any arbitrary Hilbert space; then $\Lambda_A$ is completely positive on ${\cal H}_A$ if $\Lambda_A\otimes I_B$ is positive for all such extensions. Complete positivity corresponds to the
statement that, if system $A$ evolves and system $B$ does not, any initial density matrix of the combined system evolves to another density matrix.

In trapped ions one can use the techniques of ``reservoir
engineering''~\cite{poyatos,arrc,turchette} to build different kinds of
$\hat{c}_i$ operators for the center-of-mass motion of the ion, and, in particular,
the dissipative and diffusive reservoirs discussed previously in the context
of classical dynamics. 

\subsubsection{The dissipative case}

Dissipation by a zero-temperature reservoir, in the rotating-wave approximation, is described by~(\ref{eq:lindblad}) using just one operator
$\hat{c}_1\equiv \hat a$
\begin{equation}
 \label{eq:T0}
 \frac{d\hat{\rho}}{dt}=-\frac{i}{\hbar} \left[\hat H', \hat{\rho}\right] +
 \frac{\Gamma}{2} \left(2\,\hat{a}\,\hat{\rho}\,\hat{a}^\dagger -
\hat{a}^\dagger\,\hat{a}\,\hat{\rho} -
 \hat{\rho}\,\hat{a}^\dagger\,\hat{a}\right) \,,
\end{equation}
where $\hat H'$ has the same for as $\hat H$ given by~(\ref{HAM_q}), with $\nu$ replaced by the frequency $\Omega$ given by~(\ref{Om}). We will see that, with this choice, the oscillation frequencies of the quantum and classical systems will coincide.

From the master equation~(\ref{eq:T0}), one gets the equations of motion for the expectation
values between the kicks:
\begin{eqnarray}
\langle \dot {\hat a} \rangle ={\rm Tr}\left( \hat a \dot{\hat \rho} \right) =
-i \Omega \langle \hat a \rangle -\frac{\Gamma}{2}\langle \hat a \rangle \, ,
\end{eqnarray}
which can be written in terms of $\hat x$ and $\hat p$ as 
\begin{eqnarray}
\label{eq:medxdiss}
\langle \dot {\hat x} \rangle &=& \frac{\langle \hat p \rangle}{m} -
\frac{\Gamma}{2} \langle \hat x \rangle \, , \\
\label{eq:medpdiss}
\langle \dot {\hat p} \rangle &=& -m \Omega^2\langle \hat x \rangle -
\frac{\Gamma}{2} \langle \hat p \rangle.
\end{eqnarray}

One should note that, differently from the classical equations of motion, dissipation appears here in a symmetric way with respect to position and momentum. This is related to the rotating-wave approximation, adopted in deriving~(\ref{eq:T0}): indeed, this approximation requires that the oscillator suffers many oscillation within the decay time (that is, one should have $\Gamma\ll\nu$), which implies that the effect of the dissipation gets distributed between the canonical coordinates. 

Taking the derivative of~(\ref{eq:medxdiss}) and using~(\ref{eq:medpdiss})
one gets:
\begin{equation}
\langle \ddot {\hat x} \rangle +\left(\Omega^2+\frac{\Gamma^2}{4}\right) \langle
\hat x \rangle + \Gamma \langle\dot{\hat x}\rangle = 0 \, .
\end{equation}
Using~(\ref{Om}), we can see that this equation, that describes the quantum dynamics between the kicks, is identical to its classical version~(\ref{difeqdiss}), so that the quantum and the classical systems oscillate with the same frequency.

\subsubsection{The diffusive case}

The purely diffusive reservoir master equation can obtained
from~(\ref{eq:lindblad}) by choosing two operators $\hat{c}_1=\hat a$ and
$\hat{c}_2=\hat {a}^{\dag}$ with the same rate $\gamma_1=\gamma_2=\gamma$
\begin{eqnarray}
\label{eq:medif}
 {\dot {\hat \rho}} = \frac{\gamma}{2}\Big[\left(2 {\hat a}{\hat \rho}{\hat
 a}^\dag-{\hat a}^\dag \hat a {\hat \rho}-{\hat \rho}{\hat a}^\dag \hat a
 \right) \nonumber \\
+ \left(2 {\hat{a}^\dag}{\hat \rho}{\hat a}-{\hat a} \hat{a}^\dag
 {\hat \rho}-{\hat \rho}\hat a \hat{a}^\dag \right) \Big]. 
\end{eqnarray}

This is a combination of cooling and heating reservoirs and due to the fact
that they have the same rates, all terms leading to drifts are canceled out and
only diffusion terms survive. This fact becomes clear when one writes
explicitly the corresponding Fokker-Planck equation for the Wigner function: 
\begin{equation}
\frac{\partial W}{\partial t} = \gamma \frac{\partial^2 W}{\partial \alpha
\partial \alpha^*} ,
\end{equation}
or, in terms of $\bar v$ and $\bar u$,
\begin{equation}
\frac{\partial W}{\partial t}= \frac{\gamma}{4} \left(\frac{\partial^2}{\partial \bar
v^2} + \frac{\partial^2}{\partial \bar u^2}\right)W.
\end{equation} 
This equation is equivalent to the third term of~(\ref{eq:FPcl}) rewritten in
terms of the rescaled variables if we set $\gamma={\cal D}/\eta^2$. 

A purely diffusive reservoir can be produced by random electric
fields~\cite{james,arrc} and is known to model the heating of vibrational energy observed in recent experiments on ion dynamics~\cite{turchette}.

\section{RESULTS}
\label{Results}

The classical description of a chaotic dynamical system, either using single
trajectories or a probability distribution, is based on the analysis of the
phase space and its structures. The definition of a single trajectory in the
quantum case is prevented by the uncertainty principle and the suitable
description of the system is based on quasi-probability distributions. The
Wigner function fulfills almost all the requirements for being a true
probability distribution, as it is the only quantum distribution that yields the correct marginal distributions for any direction of integration in phase space, however it can exhibit negative
values. For our purposes, this turns out to be an advantage, because
it highlights the differences between quantum and classical dynamics. As a matter of fact, it is much easier to detect quantum signatures with the Wigner distribution rather than with the Husimi or $Q$-function.

Oscillations between negative and positive values in the Wigner function are a
sign of existence of quantum interference phenomena, absent in its classical
counterpart. The role played by decoherence in washing out interference
patterns is also easily visualized in the Wigner function~\cite{habib}. More than a
visualization tool, the Wigner function can be useful to derive some
analytical results concerning the quantum-classical limit.

In what follows we make use of the Wigner function and of its Fourier
transform, the characteristic function, to obtain new results concerning the
time scales for the quantum-classical transition. Combining the interaction with the 
environment and the possibility of varying the effective Planck
constant, we are able to discuss not only the regions of parameters for the
classical limit but also the behavior of the breaking time in open
systems. 

\subsection{System without reservoir}

In the absence of the interaction with the environment, the
classical limit is investigated by changing the scaling parameter $\eta$. One should remark that, in terms of the variables $\bar u$ and $\bar v$, the initial distribution does not depend on $\eta$, and is taken to be the same for the classical and the quantum systems. Of course, in terms of the original variable $u$ and $v$, decreasing $\eta$ leads to a shrinking width of the initial distribution, both in the classical and the quantum situation. In any case, changing $\eta$ will affect both the classical and the quantum solutions, since the initial states are always taken to coincide, and a broader initial packet will explore, since the beginning, a larger region of phase space. 

\begin{figure}
\includegraphics[width=7.0cm]{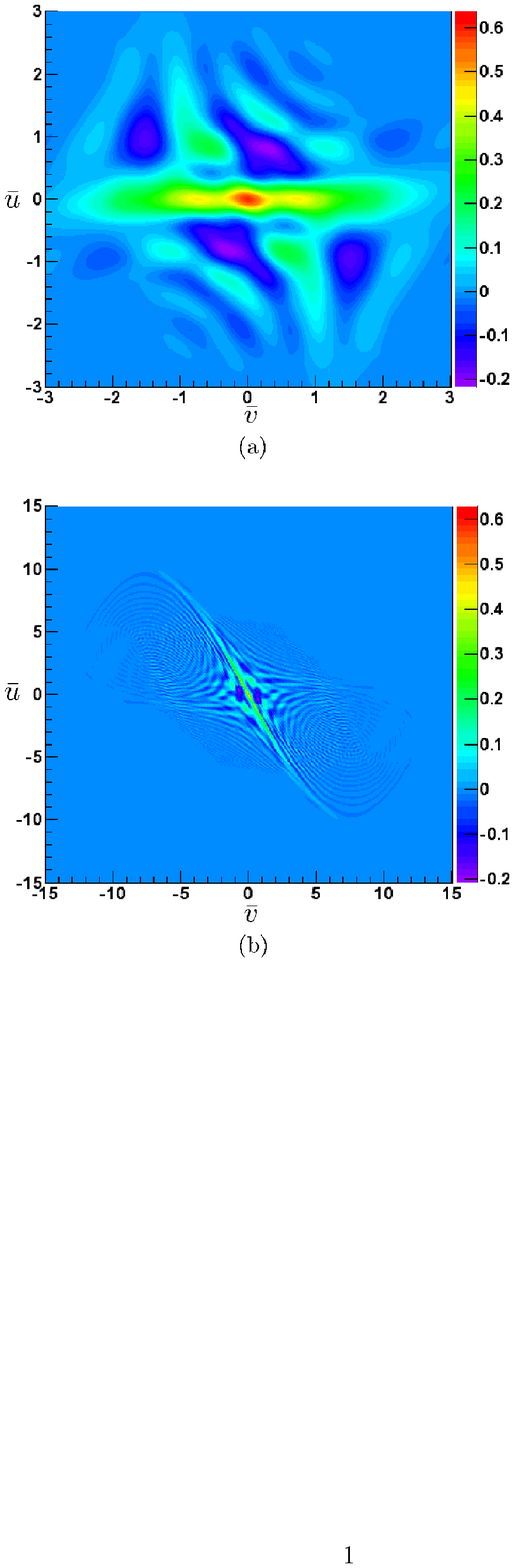}
\caption{(Color online) Wigner distribution after 9 kicks for $\eta=0.5$ (a) and $\eta=0.1$
(b). In both cases the Wigner function presents negative values but as the
Lamb-Dicke parameter is decreased (closer to the classical limit) there is a
better correspondence with the overall classical structure shown in
Fig.~\ref{figPDcons}-d.}
\label{figwigcons}
\end{figure}

In Fig.~\ref{figwigcons} we show the Wigner function for $\eta=0.5$
(top) and $\eta=0.1$ (bottom) after nine
kicks, corresponding, respectively, to the classical situation depicted in
Fig.~\ref{figPDcons}d and its scaled version (not shown). Decreasing the value
of the effective Planck constant one gets a quantum phase space that
resembles more and more the overall classical structure but still with the
presence of interference patterns.

After some time these differences between quantum and classical evolutions become important and estimates can be made using the characteristic function $C(\lambda,\lambda^*)$, defined as 
\begin{equation}
\label{fcarac}
C(\lambda,\lambda^*)={\rm Tr}\left[{\hat \rho} e^{\lambda{\hat
a}^\dag-\lambda^*\hat a}\right].
\end{equation}

Expanding the exponential of the cosine function in the quantum
map~(\ref{qmap}) in terms of Bessel functions, it is possible to write
explicitly the characteristic function after the $n$-th kick in terms of its initial value, $C_0(\lambda_n,\lambda^*_n)$, as~\cite{nonlinear4}
\begin{equation}
\label{fcaracq}
C_n(\lambda,\lambda^*)=\!\!\!\!\!\!\!\!\!\!\sum_{m_1,\ldots,m_n=-\infty}^{\infty}\!\!\!\!\!\!\!\!\!\!\!J_{m_1}(z_1)J_{m_2}(z_2)\ldots
J_{m_n}(z_n)\,C_0(\lambda_n,\lambda^*_n),
\end{equation}
where 
\begin{subequations}
\label{relations}
\begin{eqnarray}
\label{lambda}
\lambda_{k}&=&\lambda_{k-1}e^{i\alpha}+im_{k}\eta,\\
z_k&=&2K_q \,\sin(\xi_k)=\frac{K}{\eta^2} \, \sin(\xi_k), \\
\xi_k&=&-\frac{\eta}{2}(\lambda_k+\lambda_k^*), \\
\lambda_{0}&\equiv&\lambda,
\end{eqnarray}
\end{subequations}
$J_{m}$ are Bessel functions and $\alpha=\nu \tau$, as in the classical case.

It is interesting to compare this expression with the corresponding one for the classical system, which can be obtained by introducing the classical
map~(\ref{mapaescalado}) into the appropriate classical definition of the characteristic function. This definition follows from the quantum
expression~(\ref{fcarac}) upon changing the trace by a double integral, the
operators by complex numbers and the density matrix by a classical probability
density. One gets then:
\begin{eqnarray}
\label{fcaracc}
C^{\rm (cl)}_n(\lambda,\lambda^*)&=&\!\!\!\!\!\!\!\!\!\!\sum_{m_1,\ldots,m_n=-\infty}^{\infty}\!\!\!\!\!\!\!\!\!\!\!J_{m_1}\left(\frac{K}{\eta^2}\xi_1\right)\ldots
J_{m_n}\left(\frac{K}{\eta^2}\xi_n\right)\nonumber\\
&\times&C_0(\lambda_n,\lambda^*_n).
\end{eqnarray}

It is clear that the classical expression is obtained from the quantum one when $\left|\xi_k\right|\ll1$, in which case
\begin{equation}
\label{aproximacao}
\sin(\xi_k) \approx \xi_k\,.
\end{equation}
Since, according to Eq.~(\ref{lambda}), $\xi_k$ is proportional to $\eta$, this approximation should hold for sufficiently small $\eta$ in the beginning of the evolution of the system. However, as time evolves, and $|\lambda_k|$ grows, it eventually ceases to be true. This is precisely where the breakdown between classical and quantum evolution occurs. 

\begin{figure}
\includegraphics{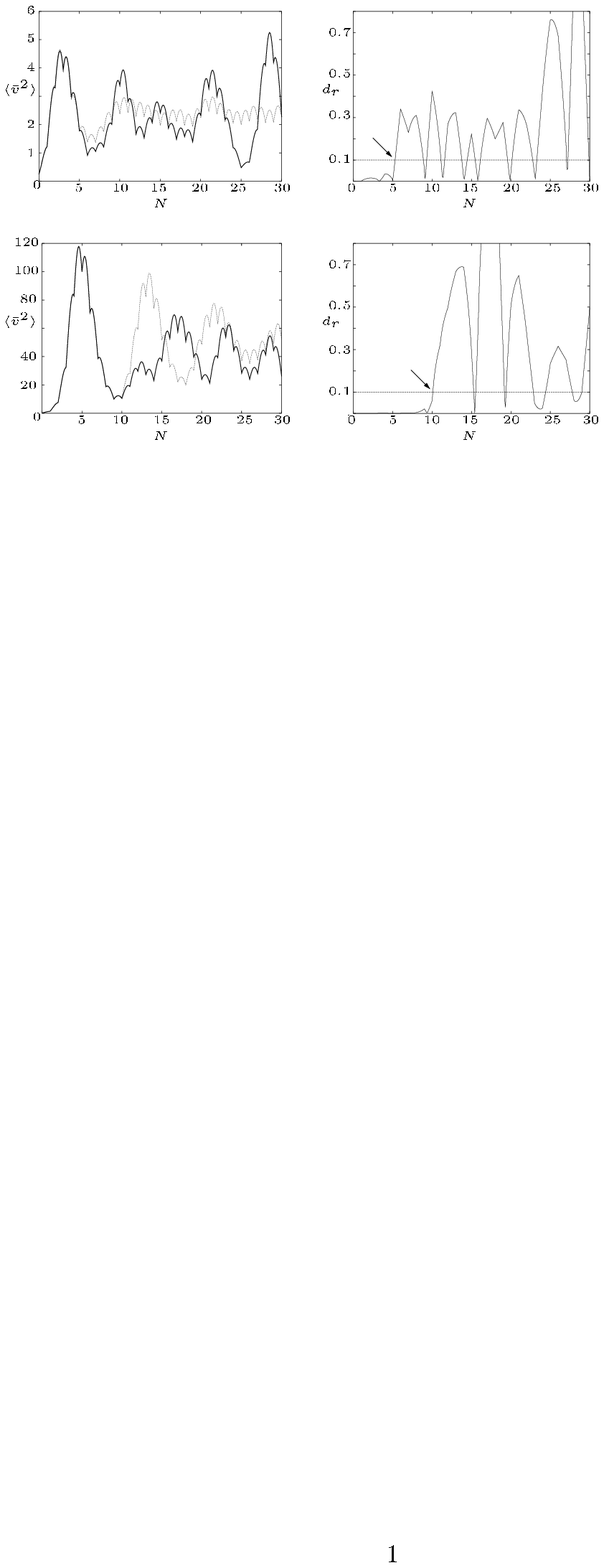}
\caption{Classical and quantum variances (left) and the relative distance
$d_r$ (right) as a function of the number of kicks for $\eta=0.5$ (top) and
$\eta=0.1$ (bottom). The quantum (solid) variance remains close to the
classical (dashed) for a longer time when the Lamb-Dicke parameter is
smaller. The breaking time, indicated by the arrows, correspond to the instant at
which the relative distance gets larger than a chosen value $\epsilon=0.1$
(horizontal line in the right panels).}
\label{figvariancias}
\end{figure}

\begin{figure}
\includegraphics{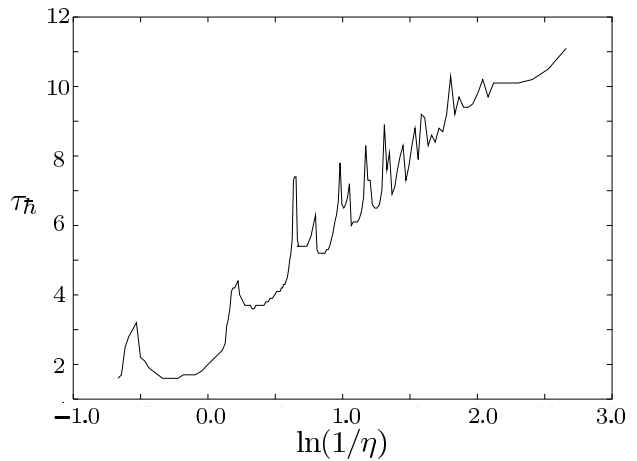}
\caption{Breaking time as a function of $\ln(1/\eta)$ for $K=2.0$ and $q=6$.
Despite the oscillations there is a clear linear behavior confirming the scaling predicted by~(\ref{tempsep}).}
\label{figtempsep}
\end{figure}

An expression for the breaking time can be obtained by comparing the quantum and classical characteristic
functions. We may define it as the time at
which the approximation~(\ref{aproximacao}) fails, or, in other words, for which 
$\xi_k\approx 1$. Assuming the strong chaos condition ($K \gg1$), one is
able to derive~\cite{nonlinear4} 
\begin{equation}
\label{tempsep}
\tau_{\hbar} \approx \frac{\ln(2 \bar K/\eta)}{\ln(\bar K)},
\end{equation}
where $\bar K=K \sin(\alpha)$. This result displays the scaling of the breaking time
 with the logarithmic of $1/\eta^2$, which stands for the $\hbar_{eff}^{-1}$ already mentioned in the Introduction. 
A numerical check of this expression needs an operational definition for the
 breaking time, which involves, also, a choice of an appropriate measure of the
 distance between quantum and classical systems. Information measures that can be used to compare two different distributions are available in the literature~\cite{beck}, and have been applied in the context of quantum-classical transition for chaotic systems~\cite{pattparameter}. Measures based on
 the comparison between the whole distributions, although more
 complete, would lead to a large increase in computation time and to
 experimental difficulties. Although the Wigner function has already been
 measured in experiments with trapped ions~\cite{wigions}, it would be
 challenging to resolve the details of the interference fringes seen,
 for example, in Fig.~\ref{figwigcons}-b. The relative distance between the
classical ($\langle\Delta \bar v_{cl}^2\rangle$) and quantum ($\langle\Delta \bar v_{q}^2\rangle$) variances of the distributions, defined as
\begin{equation}
d_r = \left|\frac{\langle\Delta \bar v_{cl}^2\rangle-\langle\Delta \bar
v_{q}^2\rangle}{\langle\Delta \bar v_{cl}^2\rangle}\right|, 
\end{equation}
is a much simpler quantity that already shows the scaling properties
of~(\ref{tempsep}). Fig.~\ref{figvariancias} shows the classical and quantum
variances (left) and the relative distance $d_r$ (right) as a function of the
number of kicks for two different Lamb-Dicke parameters. The separation time is defined as the
time at which the relative distance crosses a given value $\epsilon$
($\epsilon=0.1$ in the figure). In Fig.~\ref{figtempsep} we plot $\tau_{\hbar}$ obtained in this way as a function of $\ln(1/\eta)$ and, although the
absolute value of the breaking time depends on the choice of $\epsilon$, tests
with $\epsilon$ ranging from $5\%$ to $30\%$ show only slight modifications in
the curves and confirm the scaling behavior~(\ref{tempsep}), independently of the particular definition of the separation.

\subsection{Dissipative environment}

The analytical solution for the quantum dissipative problem in terms of the
characteristic function is also given by~(\ref{fcaracq}) and the only change
affects the relation~(\ref{lambda}) that must be replaced by
\begin{equation}
\label{lambdadiss} 
\lambda_{k}=\lambda_{k-1}e^{i{\bar \alpha}}e^{-\Gamma\tau/2}+im_{k}\eta,
\end{equation}
where $\Gamma$ was introduced in the classical case. Besides the usual rotation
due to the harmonic motion represented by the complex exponential
in~(\ref{lambdadiss}) there is also an exponential decay due to the
dissipative drift in the characteristic function's argument~\cite{mtqo}. 

In analogy with what was done for the system without reservoir, one can obtain the breaking time by examining when the quantum characteristic function cannot be described
anymore by its semiclassical approximation. This procedure is fully described
in Appendix~\ref{app:diss} and the results are summarized in
Table~\ref{table1}. The first column represents a region characterized by a
deep quantum regime where a classical description of the system is already not valid
right after the first kick, for all finite values of the dissipation. It is
interesting to note that, in this region, quantum-classical correspondence is lost even if dissipation is sufficient to bring the classical system into a periodic regime. 

In the second column we have the most interesting range of parameters concerning the quantum-classical transition (we call it ``weak quantum regime''), where two different regimes exist: one indicating
an increase of the breaking time with dissipation, and another showing close 
quantum-classical behavior for all times ($\tau_{\hbar}\rightarrow \infty$). This latter case corresponds to a situation where the dissipation is
so strong that classical chaos is suppressed and the system goes to a simple
attractor. The breaking time for region (c) in Table~\ref{table1},
\begin{equation}
\label{tempsepdiss}
\tau_{\hbar}^{dis}\approx \frac{\ln{(2\bar {K'}/\eta)}}{\ln{(\bar {K'})}-\Gamma\tau/2},
\end{equation}
with $\bar {K'}=K' \sin(\bar \alpha)$,
increases as the dissipation coefficient grows, but it keeps the
same logarithmic-scale dependence with respect to the effective Planck constant as in the
case without reservoir. 

Although $\tau_{\hbar}^{dis}$ could be arbitrarily large, as
pointed out by Iomin and Zaslavsky~\cite{iomin} in a recent derivation of an
expression similar to~(\ref{tempsepdiss}), this is not the case if one wants to
preserve a strange attractor. The condition $\Gamma\tau/2=\ln{(\bar {K'})}$, which separates
regions (c) and (d) in Table~\ref{table1}, corresponds to the
situation where the origin of phase space changes from an unstable to a stable fixed
point. However, instability of the origin is not sufficient to ensure chaotic
dynamics. Indeed, Fig.~\ref{figlyapbif} exhibits a large range of values of $\Gamma\tau/2$ 
for which the system is attracted to some periodic trajectory, even when $\Gamma\tau/2<\ln{(K)}$. 
\begin{table}
\caption{\label{table1} Breaking time for dissipative dynamics in the four different regions of parameters $\Gamma\tau/2$, $\eta$ and $\bar {K'}$.}\label{tab5.1}
\begin{ruledtabular}
\begin{tabular}{c c c}
 Nonlinearity & Deep quantum & Weak quantum \\
strength & regime & regime \\
\hline\\
 & $ \Gamma\tau/2< \ln(\eta/2)$ & $ \Gamma\tau/2> \ln(\eta/2) $ \\ 
& & \\
$\ln{(\bar {K'})}>\Gamma\tau/2$ & (a) $\tau_{\hbar}\approx 1$ kick & (c) $\tau_{\hbar}\approx \frac{\ln{(2
\bar {K'}/\eta)}}{\ln{(\bar {K'})}-\Gamma\tau/2}$ \\ 
& & \\
$\ln{(\bar {K'})}<\Gamma\tau/2$ & (b) $\tau_{\hbar}\approx 1$ kick & (d) $\tau_{\hbar}\rightarrow \infty$\\
\end{tabular}
\end{ruledtabular}
\end{table}

\begin{figure}
\includegraphics{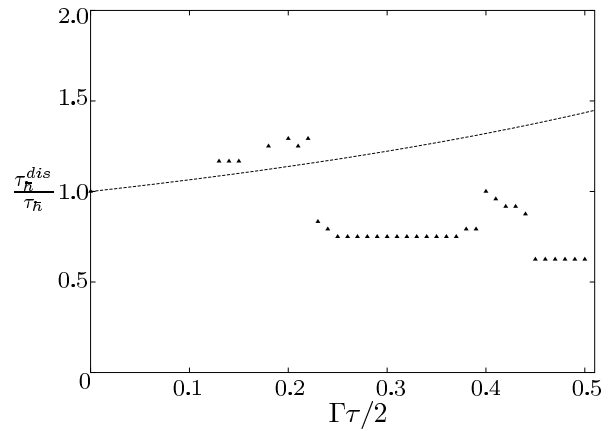}
\caption{Ratio between the breaking time with and without dissipation as a
function of $\Gamma\tau$, for the same parameters as in Fig.~\ref{figlyapbif}
and $\eta=0.5$. Solid
line shows the analytical prediction~(\ref{razaotempsep}) while triangles show
numerical simulations comparing quantum and classical systems. The differences
between quantum and classical environments are responsible for the deviations
observed for larger values of $\Gamma\tau$. Smaller values of $\Gamma\tau$ where not considered 
due to numerical constraints.}
\label{figtempsepdiss}
\end{figure}

The ratio between the breaking times with and without dissipation is given by
\begin{equation}
\label{razaotempsep}
\frac{\tau_{\hbar}^{dis}}{\tau_{\hbar}}\approx\frac{\ln{(\bar {K'})}}{\ln{(\bar {K'})}-\Gamma\tau/2}\,.
\end{equation}
This expression exhibits the increase of the breaking time as a function of
$\Gamma\tau$. In Fig.~\ref{figtempsepdiss} this relation is plotted together with
numerical simulations for $K'=6.0$ with the horizontal axis ending at the value $(\Gamma\tau/2)_{max}\approx0.51$ of the dissipation constant, for which the
Lyapunov exponent becomes negative (see Fig.~\ref{figlyapbif}). For this value
of the kicking strength the maximum increase in the breaking time is around $1.5$
and therefore does not help considerably to achieve the classical limit. This
increase depends on the values of the chaoticity parameter but even for very
large values ($K'\approx 500$) it is less than a factor of 4. 

Although expression~(\ref{razaotempsep}) is independent of the scaling
parameter $\eta$, some remarks about the role played by the macroscopic limit
in the dissipative case are needed. It should be noted that the quantum
description of a zero-temperature reservoir given by~(\ref{eq:T0}) is not
completely equivalent to the classical description based on the
map~(\ref{mapaclassicodiss}). In fact, a classical distribution submitted only to the dissipative dynamics would
contract towards the origin, while a quantum distribution would
end up in the ground state, which has a finite width due to the uncertainty
principle. This argument does not invalidate the results presented in
Table~\ref{table1}, but rather emphasize that they are valid for the
semiclassical approximation that is not equivalent to the
fully classical system based on map~(\ref{mapaclassicodiss}). 

This discussion also provides an explanation for the deviations between
numerical and analytical results shown in Fig.~\ref{figtempsepdiss}. Only in a
small range of the dissipation strength ($0.12\lesssim \Gamma\tau/2 \lesssim 0.22$) we were
able to see the expected growth in the breaking time, while for larger values of
$\Gamma\tau/2$, the effects of the difference between classical and quantum systems,
discussed above, become predominant. 

It is interesting to note the sudden growth in
the breaking time for the region corresponding to the periodic window around
$\Gamma\tau/2=0.4$ shown in Fig.~\ref{figlyapbif}. This should be expected, since the quantum and classical systems should stay together for a longer time outside the chaotic region. On the other hand, as $\Gamma\tau$ increases, inside the same region, one notices that the breaking time decreases. This is due to the fact that, for larger dissipation, the distribution shrinks at a faster rate, implying that the two distributions approach at an earlier time the region around the origin, where the uncertainty principle plays a dominant role. 

For smaller values of the dissipation parameter, the system could spread over a
large region of phase space demanding a huge amount of computational
resources. This, together with reliability problems for even smaller values of
$\Gamma\tau/2$, imposed the limit $\Gamma\tau/2\gtrsim 0.12$ for the dissipation parameter in our calculations.

\begin{figure}
\includegraphics{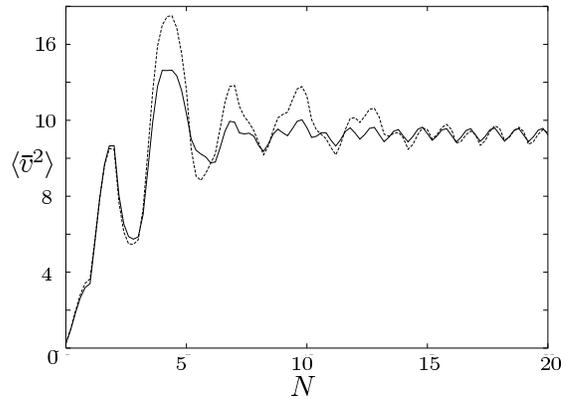}
\caption{Classical (dashed line) and quantum (solid line) evolution of
$\langle \Delta \bar v^2 \rangle$ as a function of the number of kicks $N$ for
$K'=6.0$ and $\Gamma\tau/2=0.36$. Classical and quantum variances share the same asymptotic behavior despite the fact that they can differ in a transient regime.}
\label{figdifdiss}
\end{figure}

\begin{figure}
\includegraphics[width=7.5cm]{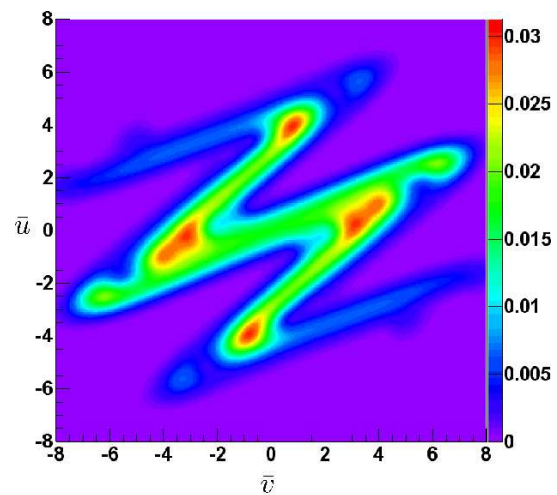}
\caption{(Color online) Wigner function for the same parameters of the strange attractor of
Fig.~\ref{figmapdiss} and $\eta=0.5$. The quantum distribution, although lying in the region
of the classical attractor, does not show the classical small-scale structures.}
\label{figwigdiss}
\end{figure}

It is important to understand the meaning of the breaking time and its
consequences for the dynamics of the system at different times. In particular,
the stationary state produced by the dissipation is of much interest and this
issue of long time behavior has been addressed before in the case of the standard map~\cite{dittrich}. The existence of a finite $\tau_{\hbar}$ means that
quantum and classical dynamics cannot be equivalent for all times but does
not necessarily mean that they have to be different for all
$t>\tau_{\hbar}$. In fact, the numerical simulations for the evolution of the
variance of the dissipative KHO show that, in some cases, quantum and
classical calculations share the same final stationary behavior but with
different transient regimes as can be seen in Fig.~\ref{figdifdiss}. One should
note, however, that this is not necessarily true for the whole phase space
distributions, which can be different, although having the same second
moments. This can be illustrated through the comparison of the Wigner function
depicted in Fig.~\ref{figwigdiss} and the strange attractor shown in
Fig.~\ref{figmapdiss}. The quantum distribution clearly does not exhibit all
the structures presented in the classical case despite the fact that it lies in
the region that contains the classical attractor. 

\subsection{Diffusive environment}

The evolution of the system under the influence of the diffusive environment
can be solved analytically in terms of the characteristic function, as
described in Appendix~\ref{app:difu}. The solution between consecutive kicks 
can be written as
\begin{equation}
\label{caracdifu} 
C(\lambda,\lambda^*,t) = C(\lambda,\lambda^*,0)e^{-\gamma\vert \lambda\vert^2 \tau}.
\end{equation}

Using this solution together with~(\ref{fcaracq}), we can establish a recurrence
relation for the characteristic function after the $n$-th kick as 
\begin{equation}
\label{fcaracdifu}
C_n(\lambda,\lambda^*)=e^{-\gamma\vert \lambda\vert^2 \tau}\sum_{m_1=-\infty}^{\infty}J_{m_1}(z_1)\,C_{n-1}(\lambda_1,\lambda^*_1),
\end{equation}
with all the variables defined as in~(\ref{relations}).

While in the dissipative case the drift effect adds an exponential factor to
the arguments of the Bessel functions, in this case, a Gaussian multiplies the whole sum. This difference is crucial for understanding the influence of this environment in restoring the quantum-classical
correspondence. First, one should note that the role played by the Lamb-Dicke
parameter in the above expressions is the same as in the case of a system without reservoir and, therefore, the macroscopic approximation would lead to the same result as before. It is clear, however, that diffusion should have an important effect on the behavior of the system. This can be seen through a more careful
analysis of expression~(\ref{fcaracdifu}). 

Assume that the quantum and classical dynamics coincide at kick $n$ and
forget, for the moment, the diffusion. As discussed before, the two dynamics
will differ as long as the approximation $\sin(\xi)\approx \xi$ fails. This
gives an estimate of the values of $\xi_1$ that lead to quantum corrections as
\begin{equation}
\vert \xi_1 \vert =\left\vert \frac{\eta}{2} (\lambda_1 +\lambda^*_1)\right\vert \gtrsim 1.
\end{equation}
This equation shows that quantum corrections are associated with large values of $\lambda$ and we can define the typical values $\lambda_T$ for
which the corrections appear as
\begin{equation}
\label{conddifu}
\vert \lambda_T \vert \equiv \vert\Re(\lambda e^{i\alpha})\vert \gtrsim \frac{2}{\eta}.
\end{equation}

The effect of diffusion is to cut off the contributions from large values of
$\lambda$, due to the Gaussian modulation in~(\ref{caracdifu}). This implies
that the values of $\lambda$ that satisfy~(\ref{conddifu}) may be attenuated
by the Gaussian pre-factor, which renders them inefficient in promoting the
quantum-classical separation.

\begin{figure}
\includegraphics[width=8.5cm]{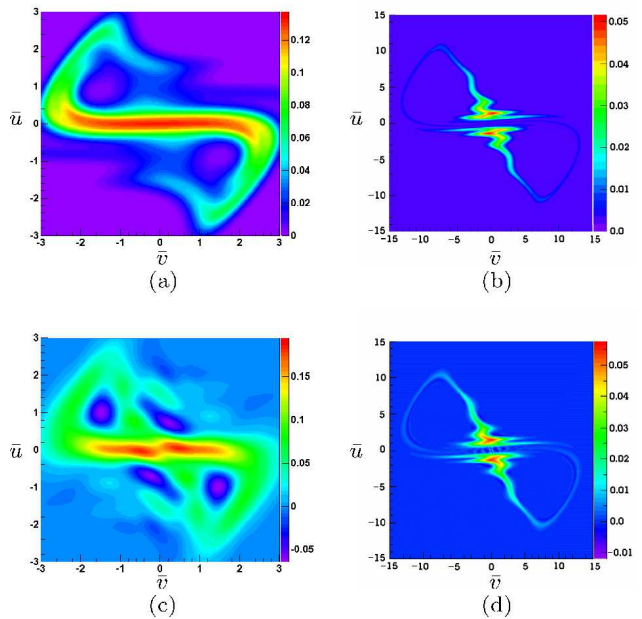}
\caption{(Color online) Classical (top) and Wigner (bottom) distributions for $K=2.0$,
$D=0.01047$ and $\eta=0.5$ (left) or $\eta=0.1$ (right). Diffusion leads to a
better quantum-classical correspondence as compared to
Fig.~\ref{figwigcons}. For $\eta=0.1$ this correspondence is quite impressive
while for $\eta=0.5$ differences still remain. Diffusion also prevents the appearance of small-scale structures on classical dynamics (top).}
\label{figwigdif}
\end{figure}

This may become more intuitive if we go back from the characteristic
to the Wigner function: because they are related by a Fourier
transform, the larger values of the characteristic function correspond to the
small-scale interference structures in the Wigner function and thus to quantum
corrections. The disappearance of these small-scale structures, due to the Gaussian modulation of the characteristic function, has been extensively discussed in the literature~\cite{kolovsky,habib,zurek,pattparameter,pattquantum}: it leads to a better correspondence between quantum and classical distributions, and to the emergence of the classical world~\cite{Zurek2}. Fig.~\ref{figwigdif} shows, for two different Lamb-Dicke
parameters, classical and Wigner distributions in the presence of
diffusion. These distributions are much more similar to each other than the corresponding distributions for the system without reservoir, displayed in Fig.~\ref{figwigcons}.
The importance of diffusive effects will depend, however, on the value of the
effective Planck constant, as can be seen by comparing the impressive
correspondence between quantum and classical distributions for $\eta=0.1$ (right) and some evident differences that
persist for $\eta=0.5$ (left). 

Although it is hard to define precisely the value of the diffusion needed to
restore the classical limit, a rough estimate can be obtained as follows: if
the values of $\lambda_T$, given by~(\ref{conddifu}) lie outside the range defined by the width of the
Gaussian, then the corrections should remain small. Using~(\ref{conddifu}), this conditions reads
\begin{eqnarray}
\frac{2}{\eta} \gtrsim \frac{1}{2\sqrt{D}},
\end{eqnarray}
where $D\equiv \gamma\tau/2$ plays the role of a dimensionless diffusion coefficient for the renormalized coordinates $(\bar u,\bar v)$. 
This simple argument shows that there is a critical diffusion for which the
classical and quantum dynamics remain close to each other, and that it scales as
\begin{equation}
\label{scaling}
D_{\rm cr} \propto \eta^2.
\end{equation}

In terms of the diffusion coefficient in Eq.~(\ref{eq:FPcl}), one has:
\begin{equation}
{\cal D}_{\rm cr}\propto\eta^4/\tau\,.
\end{equation}

This is the diffusion coefficient that corresponds to the non-renormalized variables $v$ and $u$. This result is consistent with those found for example in Refs.~\cite{kolovsky} and~\cite{pattparameter}. One should expect, however, that the strength of the nonlinearity, represented by $K$ in our case, should play an important role in such a scaling
law. The argument leading to~(\ref{scaling}) was based in the
estimation of the values of $\xi_1$ at which quantum corrections become
important, without taking into account the size of these corrections. We have not studied in detail the actual separation between the two distributions, as they evolve with time. This is the reason why our simple argument could not account for the influence
of nonlinearity, which is hidden in~(\ref{scaling}). A detailed investigation of the separation time for the diffusive case, or an estimation of the error introduced by the neglected contributions, would certainly display this dependence.  
Scaling relations between effective Planck constants, environment and
nonlinearity strengths in the context of the quantum-classical transition have
been obtained by many authors~\cite{ott,cohen,zurek,kolovsky,pattquantum} and
have motivated recent interest~\cite{pattparameter} in finding the properties of such
scaling. 

\begin{figure}
\includegraphics{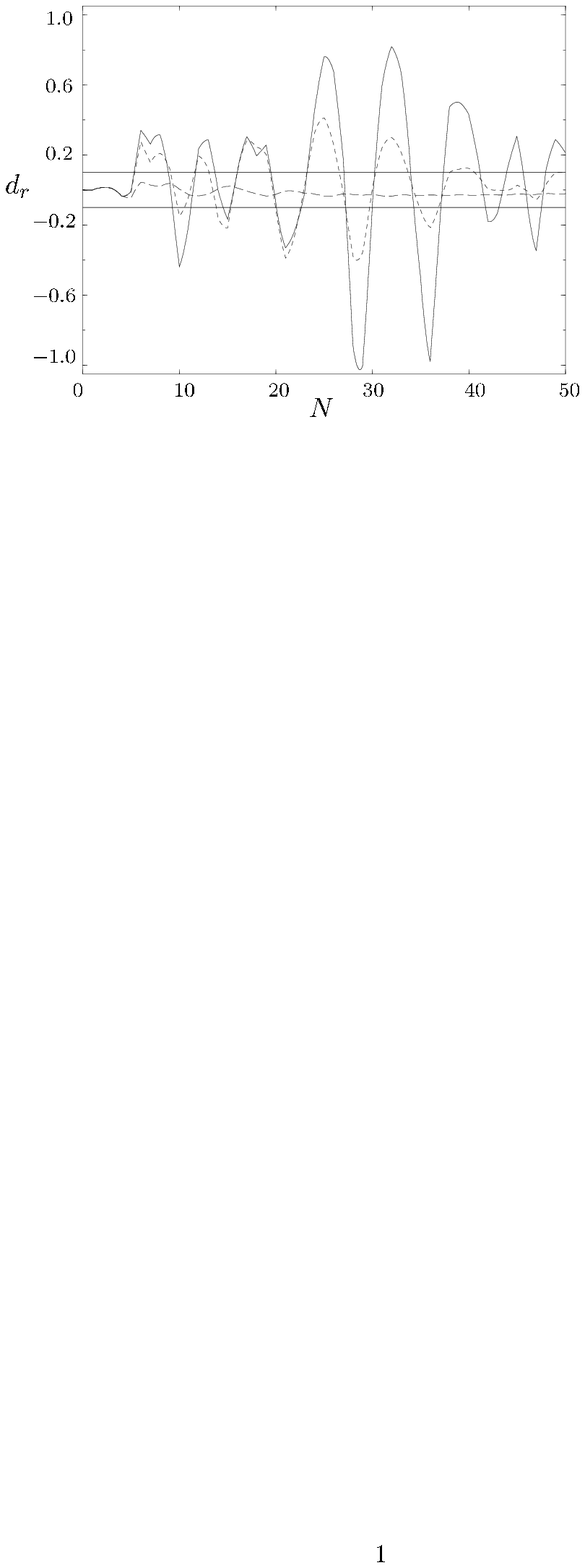}
\caption{Relative difference between quantum and classical variances as a
function of the number of kicks for $\eta=0.5$ and $D$ equal to 0 (solid);
0.00209 (dashed) and 0.0209 (long-dashed). For large enough $D$, the relative
difference remains always smaller than the percentual threshold $\epsilon$ represented by the horizontal lines.}
\label{detailsep}
\end{figure}

The above considerations suggest that the
breaking time should diverge when the diffusion coefficient exceeds a certain
critical value $D_{cr}$. This may be easily understood from Fig.~\ref{detailsep}, which displays the time evolution of the relative distance for the quantum and classical variances, for several values of $D$. One should note that, as $D$ becomes larger than a critical value, which depends on the percentual threshold $\epsilon$ adopted for the definition of the separation time, the relative distance remains always smaller than this threshold, implying an infinite separation time. On the other hand, for sufficiently small diffusion coefficients, one should recover the logarithmic time scale. Although
we were not able to derive an analytical expression for the breaking time when $D<D_{cr}$, our numerical simulations show that it remains indeed practically identical to the result obtained when no reservoir is present, in this region of parameters. 

\begin{figure}
\includegraphics{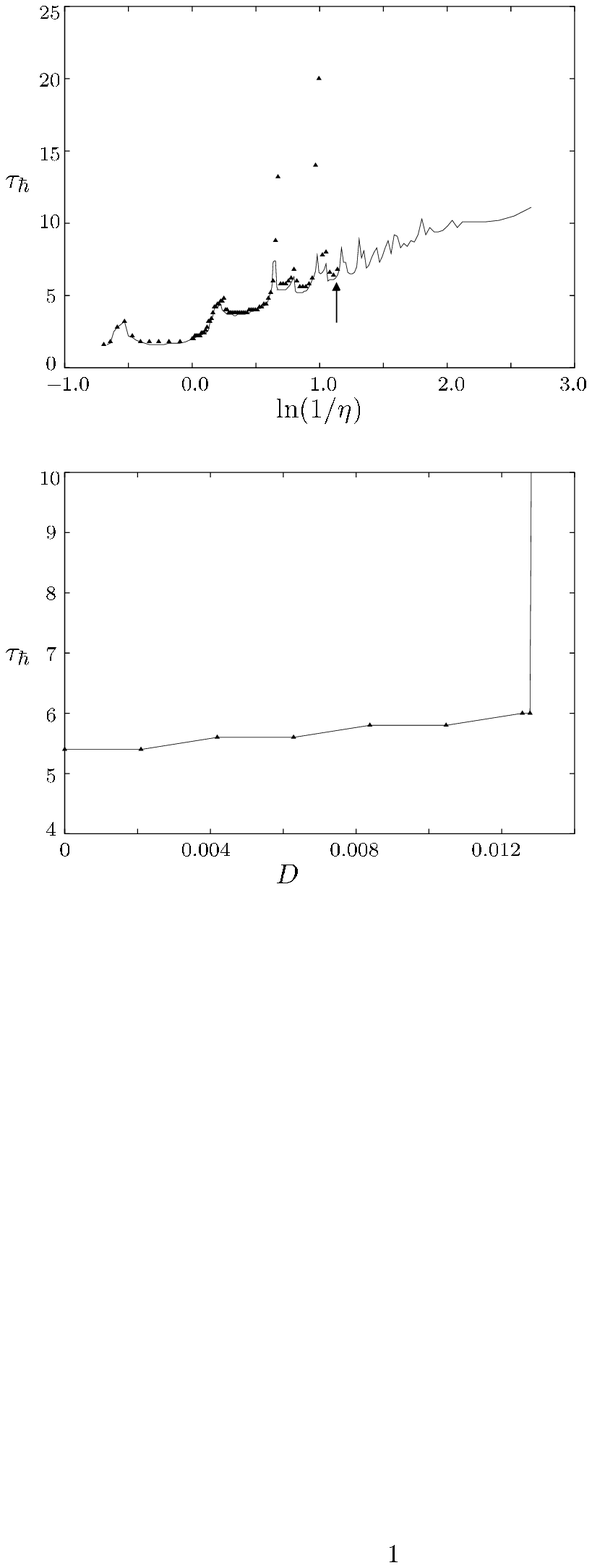}
\caption{Top: breaking time as function of $\ln(1/\eta)$ for $D=0.01047$
(triangles) and $D=0$ (line). The two curves are essentially the same until
$\eta\approx 0.31$ (shown by the arrow) when there is no separation
anymore. Some peaks of the $D=0$ curve are amplified, due, probably, to the
definition of breaking time adopted. Bottom: breaking time as function of $D$ for
$\eta=0.5$. The separation time remains basically the same,
with a small increase ($\approx 10 \, \%$ at maximum), and suddenly increases
for $D_{cr}\approx0.13$.}
\label{figdifdifu}
\end{figure}

The behavior of the breaking time in presence of a diffusive
environment is shown in Fig.~\ref{figdifdifu}. For $D=0.01047$ (top),
one can see that the breaking time basically lies on the curve corresponding to the system without reservoir, and
increases abruptly at a given value of the Lamb-Dicke parameter ($\eta=0.31$, 
indicated by an arrow in the figure), with some
of the oscillations, already present in the $D=0$ case,
amplified. For $\eta \le 0.31$, the breaking time experiences a sudden growth -- the corresponding points are not
shown in the figure (numerical tests were performed for the maximum of $50$
kicks up to $\eta=0.2$). For $\eta=0.5$ (bottom), the increase in the
noise introduces very small changes in the breaking time, which grows quickly
when $D\approx 0.01466$. Again, the abrupt increase indicates that the
differences between quantum and classical variances remain bounded below a
given limit $\epsilon$.

One should note, however, that the small changes in the breaking time for
$D<D_{cr}$ do not imply that the environment has no effect at all in the
dynamics. In fact, observing again Fig.~\ref{detailsep} we see that the
maximum distance between the variances decreases smoothly with increasing
noise strength and may eventually come to zero, indicating a perfect
quantum-classical correspondence. It is interesting to note how these two
different quantities, breaking time and maximum distance, give complementary
information about the dynamics.

From these results one could infer, naively, that diffusion is sufficient to
restore the classical limit for a chaotic system no matter the value of the
effective Planck constant. Indeed, for a given Lamb-Dicke parameter, one can
always find a large enough diffusion coefficient to bring the quantum and
classical dynamics sufficiently close to each other. Nevertheless, such a
statement deserves some reserve. In fact, diffusion washes out not only the
interference pattern in the Wigner function but also the structures in the
classical phase space (see Fig.~\ref{figwigdif}), and one could claim,
therefore, that when the critical diffusion coefficient is very large, the
chaotic characteristic of the system is lost and the system follows,
basically, a diffusive dynamics induced by the environment. A similar
situation was described in the dissipative case where a large enough dissipation was sufficient to suppress chaos and bring the system to a periodic
behavior. There, however, we could clearly distinguish between chaotic and
periodic behavior through the calculation of the Lyapunov exponent, while here,
even though generalizations of Lyapunov exponents for distributions
exist~\cite{lyapdist1,lyapdist2}, there is no sharp distinction between
chaotic and regular behavior. The description of the system can
become even more complicated with the addition of diffusion in view of the mixed phase-space structure of the system. Indeed, the difficulties in characterizing
chaos lie not only on the smoothing of the phase space structures (see
Fig.~\ref{figwigdif}-a), but also on the fact that the distribution can flow
from regular to chaotic regions that were well separated when no reservoirs were taken into account.

Finally, one should remark that we have not analyzed here the long-time behavior of the system. In particular, it is well-known that diffusion may affect quantum localization, which occurs at times much longer than those here considered~\cite{ott,dittrich,wilkinson,cohen}. In the absence of a reservoir, the long-time dynamics of the model analyzed in this paper may display either a quantum diffusion or a ballistic behavior~\cite{rebuzzini}.

\section{Conclusions}

We have shown that it is possible to discuss separately, in a physically 
relevant way, the roles of the macroscopic limit and of different
system-environment interactions in the quantum-classical transition for a
chaotic system. We have considered the kicked harmonic oscillator coupled to
two distinct reservoirs, giving rise in the classical limit to either pure dissipation (zero-temperature reservoir) or pure diffusion (random force), in a situation that could
be implemented in state-of-the-art ion trap experiments. 

In the chaotic regime, when the interaction with a reservoir is not taken into account, the classical and quantum dynamics start
diverging after a time that depends logarithmically on the ratio between 
a typical action of the system and Planck constant. We have used an
operational definition of the breaking time in terms of measurable quantities, 
which would allow an experimental test of this logarithmic time scale. 

In the dissipative case we established regions of parameters corresponding to
different time scales. There is a region where quantum corrections appear right after the first kick and quantum-classical correspondence is lost already in the
beginning of the evolution. Decreasing the Lamb-Dicke parameter, one reaches a
region where quantum-classical correspondence persists for a time that, as in the 
system without reservoir, grows only logarithmically with the classicality parameter. We
have also shown that, for a fixed effective Planck constant, close 
agreement between quantum and classical predictions is only possible for
dissipation strengths large enough to bring the system into regular behavior.

In the diffusive case, we were able to establish that the behavior of the quantum-classical separation should be markedly different, depending on whether the diffusion coefficient is above or below a certain critical value $D_{cr}$. For $D>D_{cr}$, this separation should remain small, and infinite separation times may even be obtained, at values of $D$ that depend on the definition adopted for the critical percentual separation. We have also presented numerical evidence that, for diffusion coefficients below this limit, the breaking time behaves as in the case without reservoir. Furthermore, we obtained an analytical estimation of the dependence of the critical diffusion coefficient on the effective Planck
constant, which shows that the farther away from the classical limit is the system,
the larger must be the effect of the environment to restore the
quantum-classical correspondence. 

Although the coupling with the environment helps restore the quantum-classical correspondence for a system that 
is close to the macroscopic regime, for systems in a deep quantum region the
critical diffusion can be so large that it brings classicality at the expense
of reducing, or even extinguishing, the chaotic features of the system.

The behavior of the system under the influence of other kinds of environment
could also be explored in this context. Thermal and phase reservoirs are
examples of different environments, already produced in ion trap 
experiments, that could be used for this purpose, and could lead to 
interesting results for the quantum-classical transition scenario.

\begin{acknowledgments}
A.R.R.C. thanks A. Buchleitner and S. Wimberger for many useful discussions
and H. H\"affner for comments on experimental issues. This work was partially
supported by PRONEX (Programa de Apoio a N\'ucleos de Excel\^encia), CNPq
(Conselho Nacional de Desenvolvimento Cient\'\i fico e Tecnol\'ogico), FAPERJ
(Funda\c c\~ao de Amparo \`a Pesquisa do Estado do Rio de Janeiro), FUJB
(Funda\c c\~ao Universit\'aria Jos\'e Bonif\'acio), and the Brazilian Millennium Institute on Quantum Information.
\end{acknowledgments}

\appendix
\section{\label{app:diss} Breaking time}

It is not difficult to generalize the solutions~(\ref{fcaracq})
and~(\ref{relations}), which were obtained for the characteristic function
corresponding to a system without reservoir, to the dissipative regime. Between two consecutive kicks the harmonic evolution
just rotates the system in phase space, and this effect appears only in the complex
exponential in~(\ref{lambda}). The solution of the dissipative master equation
simply adds an exponential decay $e^{-\Gamma \tau/2}$~\cite{mtqo} and,
therefore, the solution of the full problem is still given by~(\ref{fcaracq}):
\begin{equation}
\label{fcaracqa}
C_n(\lambda,\lambda^*)=\!\!\!\!\!\!\!\!\!\!\sum_{m_1,\ldots,m_n=-\infty}^{\infty}\!\!\!\!\!\!\!\!\!\!\!J_{m_1}(z_1)J_{m_2}(z_2)\ldots
J_{m_n}(z_n)\,C_0(\lambda_n,\lambda^*_n),
\end{equation}
where
\begin{subequations}
\label{relationsdiss}
\begin{eqnarray}
\lambda_{k}&=&\lambda_{k-1}e^{i{\bar \alpha}}e^{-\Gamma \tau/2}+i2m_{k}\eta,\\
\label{relationsdissb}
z_k&=&2K_q \,\sin(\xi_k)=\frac{K'}{\eta^2} \, \sin(\xi_k), \\
\xi_k&=&-\frac{\eta}{2}(\lambda_k+\lambda_k^*), \\
\lambda_{0}&\equiv&\lambda.
\end{eqnarray}
\end{subequations}

The procedure to obtain the breaking time in the
dissipative case follows very closely the one used in~\cite{nonlinear4} for the
situation without reservoir. First we should note that the macroscopic limit, as
discussed previously, is achieved by letting $\eta\rightarrow 0$, 
which means that the sine functions in~(\ref{relationsdissb}) can be
approximated by their argument, i.e.,
\begin{equation}
\label{approxsine}
\sin(\xi_k)\approx \xi_k.
\end{equation}
In this limit, we obtain the semiclassical characteristic function
\begin{eqnarray}
\label{fcaraccl}
C_n(\lambda,\lambda^*)&=&\sum_{m_1,\ldots,m_n=-\infty}^{\infty}J_{m_1}(2K_q
\,\xi_1)\ldots J_{m_n}(2K_q \,\xi_n)\nonumber\\
&\times&C_0(\lambda_n,\lambda^*_n),
\end{eqnarray}
with initial condition
\begin{equation}
\label{fcarac0}
C_0^{cl}(\lambda,\lambda^*)=\int_{-\infty}^{\infty} d^2\mu \,
P_0(\mu,\mu^*)e^{\lambda_n\mu^*-\lambda_n^*\mu}\, .
\end{equation}
When no reservoir is present, the semiclassical approximation leads to
the classical characteristic function derived, directly, through the
corresponding classical map. Here this is not the case, due to the fact that
the zero-temperature reservoir leads to distinct features in the quantum and the 
classical models. If submitted only to the dissipative dynamics, the quantum
system will end up in the its ground state, which has a finite width, while a classical
probability distribution would contract to a point located at the origin. In
this way, the semiclassical characteristic function shows exactly the same
nonlinear dynamics as the classical but with an intrinsic quantum property due
to the uncertainty principle. Of course, this effect becomes smaller as quantum
fluctuations become negligible compared to the size of the system, which occurs
for small values of the Lamb-Dicke parameter.

The replacement of~(\ref{fcaracq}) by~(\ref{fcaraccl}) is only valid
if~(\ref{approxsine}) holds for every $k$. Taking into account that the Bessel
functions decrease exponentially for $\vert m_k \vert \gg 2K_q \,\xi_k$, we can
truncate the sums in~(\ref{fcaracqa}), by estimating the maximum values of 
$\vert m_k \vert$ in each sum. The relevant contributions are the ones in which
the Bessel function index is of the order of its argument and, therefore,
\begin{widetext}
\begin{eqnarray}
\label{mtruncdiss}
\vert m_1\vert &\approx& 2K_q\vert \xi_1\vert=\frac{K'}{2
\eta}e^{-\Gamma \tau /2}\Big\vert\lambda e^{i{\bar \alpha}}+ \lambda^*e^{-i{\bar \alpha}}\Big\vert,
\nonumber\\
\vert m_2\vert &\approx& 2K_q\vert \xi_2\vert=\frac{K'}{2\eta}\Big\vert
(\lambda e^{i2{\bar \alpha}}+ \lambda^* e^{-i2{\bar \alpha}})e^{-2\Gamma \tau /2}-2m_1\eta\,\sin({\bar \alpha}) e^{-\Gamma \tau /2}\Big\vert,\nonumber\\
\vert m_n \vert &\approx& 2K_q\vert \xi_n\vert=\frac{K'}{2\eta}\Big\vert (\lambda e^{in{\bar \alpha}}+ \lambda^*
e^{-in{\bar \alpha}})e^{-n\Gamma \tau /2}-2m_1\eta\,\sin[(n-1){\bar \alpha}]e^{-(n-1)\Gamma \tau /2}- \ldots
-2m_{n-1}\eta\,\sin ({\bar \alpha})e^{-\Gamma \tau /2}\Big\vert\nonumber.
\end{eqnarray}
\end{widetext}
Considering, now, the strong chaos limit ($K' \gg 1$), we have
\begin{eqnarray}
\vert m_1 \vert \!\! \approx \frac{\bar {K'}}{2 \eta\,\sin({\bar \alpha})}e^{-\Gamma \tau /2}, \ldots, \vert
m_n \vert \! \approx \frac{{\bar {K'}}^n}{2 \eta\,\sin({\bar \alpha})}e^{-n\Gamma \tau /2} \nonumber.
\end{eqnarray}

From the above considerations we can study the regime of validity of the
approximation $\vert \xi_k\vert \ll 1$ for all $k$. The ratio between two consecutive $\xi$'s is given by
\begin{eqnarray}
\label{condrazao}
\frac{\vert \xi_k \vert}{\vert \xi_{k-1} \vert}\approx \bar {K'}\,e^{-\Gamma \tau /2},
\end{eqnarray}
while the first argument is
\begin{eqnarray}
\label{condprim}
\vert \xi_1 \vert \approx \frac{\eta e^{-\Gamma \tau /2}}{2}.
\end{eqnarray}

From~(\ref{condprim}) we can get two different conditions on the first
argument $\vert \xi_1\vert$. If $\Gamma \tau /2< \ln(\eta/2)$ then $\vert \xi_1 \vert > 1$
 and, right after the first kick, the classical approximation is
not valid anymore, so quantum predictions should differ from the classical ones. This
corresponds to the results shown in the first column of Table~(\ref{table1}), 
which are represented by $\tau_{\hbar}^{dis}\approx 1$ kick.
If, on the other hand, $\Gamma \tau /2 > \ln(\eta/2)$ and, therefore, $\vert \xi_1
\vert < 1$, then we have two new possibilities depending on condition~(\ref{condrazao}). 

Choosing $\Gamma \tau /2 > \ln({\bar {K'}})$, the ratio between two consecutive $\xi$'s is 
less than one and the sequence of $\xi_k$'s is decreasing with increasing
$k$. Because $\vert \xi_1\vert < 1$, all the terms will be smaller than one
and quantum and classical evolutions should stay close to each other at all
times ($\tau_{\hbar}^{dis} \rightarrow \infty$). Nevertheless, if we have $\Gamma \tau /2
< \ln({\bar {K'}})$ the sequence of $\xi_k$'s is increasing and one should expect that
there is a $\xi_k$ for which the condition $\vert \xi_k \vert < 1$ is not
fulfilled anymore and a breaking time will exist. This will happen when
$\vert \xi_k \vert \approx 1$, which can be expressed as 
\begin{eqnarray}
\frac{{\bar {K'}}^n \eta e^{-n\Gamma \tau /2}}{2\bar {K'}} \approx 1.
\end{eqnarray}
Taking the logarithm and noting that $n$, the number of kicks, corresponds to
the time in units of $\tau$, we arrive at an estimate for the breaking time as
\begin{equation}
\tau_{\hbar}^{dis} \equiv n \approx \frac{\ln(2 \bar {K'}/\eta)}{\ln{(\bar {K'})}-\Gamma \tau /2}.
\end{equation}

\section{\label{app:difu} Diffusive case}

The evolution of the symmetric-ordered characteristic function defined
in~(\ref{fcarac}) is given by
\begin{equation}
\label{fcaracevol}
 \dot C(\lambda,\lambda^*)={\rm Tr}\left[\dot{\hat \rho} e^{\lambda {{\hat
 a}^\dag} -\lambda^*\hat a}\right].
\end{equation}

The diffusive dynamics is now introduced by the replacement of $\dot{\hat
\rho}$ in the above equation by~(\ref{eq:medif}), which gives
\begin{eqnarray}
\dot C(\lambda,\lambda^*)=\frac{\gamma}{2}{\rm Tr}\Bigl[-{\hat a}^\dag
\hat a \rho e^{\lambda {{\hat a}^\dag} -\lambda^*\hat a} +2{\hat a}\rho{\hat a}^\dag
e^{\lambda {{\hat a}^\dag} -\lambda^*\hat a} \nonumber \\ -\rho{\hat a}^\dag
\hat a e^{\lambda {{\hat a}^\dag} -\lambda^*\hat a}
-\hat a {\hat a}^\dag \rho e^{\lambda {{\hat a}^\dag}-\lambda^*\hat a}
\nonumber \\+2{\hat a}^\dag \rho \hat a e^{\lambda {{\hat a}^\dag} -\lambda^*\hat a} -
\rho{\hat a}{\hat a}^\dag e^{\lambda {{\hat a}^\dag}-\lambda^*\hat a}\Bigr].
\end{eqnarray}
Rewriting the exponentials using the Baker-Hausdorff formula and the ordering properties
\begin{subequations}
\begin{eqnarray}
e^{-\beta {{\hat a}^\dag}}f(\hat a,{{\hat a}^\dag}) e^{\beta {{\hat a}^\dag}}&=&f(\hat a+\beta,{{\hat a}^\dag})\,, \\
e^{\beta \hat a}f(\hat a,{{\hat a}^\dag}) e^{-\beta \hat a}&=&f(\hat a,{{\hat a}^\dag}+\beta),
\end{eqnarray}
\end{subequations}
we obtain
\begin{eqnarray}
\dot C(\lambda,\lambda^*)&=&\frac{\gamma}{2}{\rm Tr}\left[-2\vert
 \lambda \vert^2 e^{\lambda {{\hat a}^\dag}-\lambda^*\hat a}{\hat \rho} \right]
 \nonumber \\
&=& -\Gamma \vert
 \lambda \vert^2 C(\lambda,\lambda^*).
\end{eqnarray}

This equation can be readily integrated, giving as solution
\begin{equation}
C(\lambda,\lambda^*,t)=C(\lambda,\lambda^*,0)e^{-\gamma \vert \lambda
 \vert^2 t}.
\end{equation}

\bibliography{kho}

\begin{thebibliography}{40}
\expandafter\ifx\csname natexlab\endcsname\relax\def\natexlab#1{#1}\fi
\expandafter\ifx\csname bibnamefont\endcsname\relax
  \def\bibnamefont#1{#1}\fi
\expandafter\ifx\csname bibfnamefont\endcsname\relax
  \def\bibfnamefont#1{#1}\fi
\expandafter\ifx\csname citenamefont\endcsname\relax
  \def\citenamefont#1{#1}\fi
\expandafter\ifx\csname url\endcsname\relax
  \def\url#1{\texttt{#1}}\fi
\expandafter\ifx\csname urlprefix\endcsname\relax\def\urlprefix{URL }\fi
\providecommand{\bibinfo}[2]{#2}
\providecommand{\eprint}[2][]{\url{#2}}

\bibitem[{\citenamefont{Berman and Zaslavsky}(1978)}]{log}
\bibinfo{author}{\bibfnamefont{G.~P.} \bibnamefont{Berman}} \bibnamefont{and}
  \bibinfo{author}{\bibfnamefont{G.~M.} \bibnamefont{Zaslavsky}},
  \bibinfo{journal}{Physica A} \textbf{\bibinfo{volume}{91}},
  \bibinfo{pages}{450} (\bibinfo{year}{1978}).

\bibitem[{\citenamefont{Zurek}(1998)}]{Hyperion}
\bibinfo{author}{\bibfnamefont{W.~H.} \bibnamefont{Zurek}},
  \bibinfo{journal}{Acta Phys. Polon.} \textbf{\bibinfo{volume}{B29}},
  \bibinfo{pages}{3689} (\bibinfo{year}{1998}).

\bibitem[{\citenamefont{Berry}(2001)}]{Hyperion3}
\bibinfo{author}{\bibfnamefont{M.}~\bibnamefont{Berry}}, in
  \emph{\bibinfo{booktitle}{Quantum mechanics: Scientific perpectives on Divine
  Action}}, edited by \bibinfo{editor}{\bibfnamefont{K.~W.-M.}
  \bibnamefont{Robert John~Russell}, \bibfnamefont{Philip~Clayton}}
  \bibnamefont{and}
  \bibinfo{editor}{\bibfnamefont{J.}~\bibnamefont{Polkinghorne}}
  (\bibinfo{publisher}{Vatican Observatory - CTNS Publications},
  \bibinfo{year}{2001}), pp. \bibinfo{pages}{41--54}.

\bibitem[{\citenamefont{Feynman and Vernon}(1963)}]{feynman}
\bibinfo{author}{\bibfnamefont{R.~P.} \bibnamefont{Feynman}} \bibnamefont{and}
  \bibinfo{author}{\bibfnamefont{F.~L.} \bibnamefont{Vernon}},
  \bibinfo{journal}{Ann. Phys.} \textbf{\bibinfo{volume}{24}},
  \bibinfo{pages}{118} (\bibinfo{year}{1963}).

\bibitem[{\citenamefont{Caldeira and Leggett}(1983{\natexlab{a}})}]{caldeira1}
\bibinfo{author}{\bibfnamefont{A.~O.} \bibnamefont{Caldeira}} \bibnamefont{and}
  \bibinfo{author}{\bibfnamefont{A.~J.} \bibnamefont{Leggett}},
  \bibinfo{journal}{Physica A} \textbf{\bibinfo{volume}{121}},
  \bibinfo{pages}{587} (\bibinfo{year}{1983}{\natexlab{a}}).

\bibitem[{\citenamefont{Caldeira and Leggett}(1983{\natexlab{b}})}]{caldeira2}
\bibinfo{author}{\bibfnamefont{A.~O.} \bibnamefont{Caldeira}} \bibnamefont{and}
  \bibinfo{author}{\bibfnamefont{A.~J.} \bibnamefont{Leggett}},
  \bibinfo{journal}{Ann. Phys.} \textbf{\bibinfo{volume}{149}},
  \bibinfo{pages}{374} (\bibinfo{year}{1983}{\natexlab{b}}).

\bibitem[{\citenamefont{Ott et~al.}(1984)\citenamefont{Ott, Antonsen, and
  Hanson}}]{ott}
\bibinfo{author}{\bibfnamefont{E.}~\bibnamefont{Ott}},
  \bibinfo{author}{\bibfnamefont{T.~M.} \bibnamefont{Antonsen}},
  \bibnamefont{and} \bibinfo{author}{\bibfnamefont{J.~D.}
  \bibnamefont{Hanson}}, \bibinfo{journal}{Phys. Rev. Lett.}
  \textbf{\bibinfo{volume}{53}}, \bibinfo{pages}{2187} (\bibinfo{year}{1984}).

\bibitem[{\citenamefont{Dittrich and Graham}(1990)}]{dittrich}
\bibinfo{author}{\bibfnamefont{T.}~\bibnamefont{Dittrich}} \bibnamefont{and}
  \bibinfo{author}{\bibfnamefont{R.}~\bibnamefont{Graham}},
  \bibinfo{journal}{Ann. Phys.} \textbf{\bibinfo{volume}{200}},
  \bibinfo{pages}{363} (\bibinfo{year}{1990}).

\bibitem[{\citenamefont{Wilkinson and Austin}(1992)}]{wilkinson}
\bibinfo{author}{\bibfnamefont{M.}~\bibnamefont{Wilkinson}} \bibnamefont{and}
  \bibinfo{author}{\bibfnamefont{E.~J.} \bibnamefont{Austin}},
  \bibinfo{journal}{Phys. Rev. A} \textbf{\bibinfo{volume}{46}},
  \bibinfo{pages}{64} (\bibinfo{year}{1992}).

\bibitem[{\citenamefont{Cohen}(1994)}]{cohen}
\bibinfo{author}{\bibfnamefont{D.}~\bibnamefont{Cohen}}, \bibinfo{journal}{J.
  Phys. A} \textbf{\bibinfo{volume}{27}}, \bibinfo{pages}{4805}
  (\bibinfo{year}{1994}).

\bibitem[{\citenamefont{Kolovsky}(1996)}]{kolovsky}
\bibinfo{author}{\bibfnamefont{A.~R.} \bibnamefont{Kolovsky}},
  \bibinfo{journal}{Phys. Rev. Lett.} \textbf{\bibinfo{volume}{76}},
  \bibinfo{pages}{340} (\bibinfo{year}{1996}).

\bibitem[{\citenamefont{Zurek and Paz}(1994)}]{zurek}
\bibinfo{author}{\bibfnamefont{W.~H.} \bibnamefont{Zurek}} \bibnamefont{and}
  \bibinfo{author}{\bibfnamefont{J.~P.} \bibnamefont{Paz}},
  \bibinfo{journal}{Phys. Rev. Lett.} \textbf{\bibinfo{volume}{72}},
  \bibinfo{pages}{2508} (\bibinfo{year}{1994}).

\bibitem[{\citenamefont{Pattanayak}(1999)}]{pattquantum}
\bibinfo{author}{\bibfnamefont{A.~K.} \bibnamefont{Pattanayak}},
  \bibinfo{journal}{Phys. Rev. Lett.} \textbf{\bibinfo{volume}{83}},
  \bibinfo{pages}{4526} (\bibinfo{year}{1999}).

\bibitem[{\citenamefont{Pattanayak et~al.}(2003)\citenamefont{Pattanayak,
  Sundaram, and Greenbaum}}]{pattparameter}
\bibinfo{author}{\bibfnamefont{A.~K.} \bibnamefont{Pattanayak}},
  \bibinfo{author}{\bibfnamefont{B.}~\bibnamefont{Sundaram}}, \bibnamefont{and}
  \bibinfo{author}{\bibfnamefont{B.~D.} \bibnamefont{Greenbaum}},
  \bibinfo{journal}{Phys. Rev. Lett.} \textbf{\bibinfo{volume}{90}},
  \bibinfo{pages}{014103} (\bibinfo{year}{2003}).

\bibitem[{\citenamefont{Habib et~al.}(1998)\citenamefont{Habib, Shizume, and
  Zurek}}]{habib}
\bibinfo{author}{\bibfnamefont{S.}~\bibnamefont{Habib}},
  \bibinfo{author}{\bibfnamefont{K.}~\bibnamefont{Shizume}}, \bibnamefont{and}
  \bibinfo{author}{\bibfnamefont{W.~H.} \bibnamefont{Zurek}},
  \bibinfo{journal}{Phys. Rev. Lett.} \textbf{\bibinfo{volume}{80}},
  \bibinfo{pages}{4361} (\bibinfo{year}{1998}).

\bibitem[{\citenamefont{Borgonovi and Rebuzzini}(1995)}]{rebuzzini}
\bibinfo{author}{\bibfnamefont{F.}~\bibnamefont{Borgonovi}} \bibnamefont{and}
  \bibinfo{author}{\bibfnamefont{L.}~\bibnamefont{Rebuzzini}},
  \bibinfo{journal}{Phys. Rev. E} \textbf{\bibinfo{volume}{52}},
  \bibinfo{pages}{2302} (\bibinfo{year}{1995}).

\bibitem[{\citenamefont{Zaslavsky et~al.}(1992)\citenamefont{Zaslavsky,
  Sagdeev, Usikov, and Chernikov}}]{weakchaos}
\bibinfo{author}{\bibfnamefont{G.~M.} \bibnamefont{Zaslavsky}},
  \bibinfo{author}{\bibfnamefont{R.~Z.} \bibnamefont{Sagdeev}},
  \bibinfo{author}{\bibfnamefont{D.~A.} \bibnamefont{Usikov}},
  \bibnamefont{and} \bibinfo{author}{\bibfnamefont{A.~A.}
  \bibnamefont{Chernikov}}, \emph{\bibinfo{title}{Weak chaos and quasi-regular
  patterns}} (\bibinfo{publisher}{Cambridge University Press},
  \bibinfo{year}{1992}).

\bibitem[{\citenamefont{Berman et~al.}(1991)\citenamefont{Berman, Rubaev, and
  Zaslavsky}}]{nonlinear4}
\bibinfo{author}{\bibfnamefont{G.~P.} \bibnamefont{Berman}},
  \bibinfo{author}{\bibfnamefont{V.~Y.} \bibnamefont{Rubaev}},
  \bibnamefont{and} \bibinfo{author}{\bibfnamefont{G.~M.}
  \bibnamefont{Zaslavsky}}, \bibinfo{journal}{Nonlinearity}
  \textbf{\bibinfo{volume}{4}}, \bibinfo{pages}{543} (\bibinfo{year}{1991}).

\bibitem[{\citenamefont{Hu et~al.}(1998)\citenamefont{Hu, Li, Liu, and
  Zhou}}]{bambi}
\bibinfo{author}{\bibfnamefont{B.}~\bibnamefont{Hu}},
  \bibinfo{author}{\bibfnamefont{B.}~\bibnamefont{Li}},
  \bibinfo{author}{\bibfnamefont{J.}~\bibnamefont{Liu}}, \bibnamefont{and}
  \bibinfo{author}{\bibfnamefont{J.~L.} \bibnamefont{Zhou}},
  \bibinfo{journal}{Phys. Rev. E} \textbf{\bibinfo{volume}{58}},
  \bibinfo{pages}{1743} (\bibinfo{year}{1998}).

\bibitem[{\citenamefont{Gardiner et~al.}(1997)\citenamefont{Gardiner, Cirac,
  and Zoller}}]{gardiner}
\bibinfo{author}{\bibfnamefont{S.~A.} \bibnamefont{Gardiner}},
  \bibinfo{author}{\bibfnamefont{J.~I.} \bibnamefont{Cirac}}, \bibnamefont{and}
  \bibinfo{author}{\bibfnamefont{P.}~\bibnamefont{Zoller}},
  \bibinfo{journal}{Phys. Rev. Lett.} \textbf{\bibinfo{volume}{79}},
  \bibinfo{pages}{4790} (\bibinfo{year}{1997}).

\bibitem[{\citenamefont{Poyatos et~al.}(1996)\citenamefont{Poyatos, Cirac, and
  Zoller}}]{poyatos}
\bibinfo{author}{\bibfnamefont{J.~F.} \bibnamefont{Poyatos}},
  \bibinfo{author}{\bibfnamefont{J.~I.} \bibnamefont{Cirac}}, \bibnamefont{and}
  \bibinfo{author}{\bibfnamefont{P.}~\bibnamefont{Zoller}},
  \bibinfo{journal}{Phys. Rev. Lett.} \textbf{\bibinfo{volume}{77}},
  \bibinfo{pages}{4728} (\bibinfo{year}{1996}).

\bibitem[{\citenamefont{Turchette et~al.}(2000)\citenamefont{Turchette, Myatt,
  King, Sackett, Kielpinski, Itano, Monroe, and Wineland}}]{turchette}
\bibinfo{author}{\bibfnamefont{Q.~A.} \bibnamefont{Turchette}},
  \bibinfo{author}{\bibfnamefont{C.~J.} \bibnamefont{Myatt}},
  \bibinfo{author}{\bibfnamefont{B.~E.} \bibnamefont{King}},
  \bibinfo{author}{\bibfnamefont{C.~A.} \bibnamefont{Sackett}},
  \bibinfo{author}{\bibfnamefont{D.}~\bibnamefont{Kielpinski}},
  \bibinfo{author}{\bibfnamefont{W.~M.} \bibnamefont{Itano}},
  \bibinfo{author}{\bibfnamefont{C.}~\bibnamefont{Monroe}}, \bibnamefont{and}
  \bibinfo{author}{\bibfnamefont{D.~J.} \bibnamefont{Wineland}},
  \bibinfo{journal}{Phys. Rev. A} \textbf{\bibinfo{volume}{62}},
  \bibinfo{pages}{053807} (\bibinfo{year}{2000}).

\bibitem[{\citenamefont{Iomin and Zaslavsky}(2003)}]{iomin}
\bibinfo{author}{\bibfnamefont{A.}~\bibnamefont{Iomin}} \bibnamefont{and}
  \bibinfo{author}{\bibfnamefont{G.~M.} \bibnamefont{Zaslavsky}},
  \bibinfo{journal}{Phys. Rev. E.} \textbf{\bibinfo{volume}{67}},
  \bibinfo{pages}{027203} (\bibinfo{year}{2003}).

\bibitem[{\citenamefont{Chernikov et~al.}(1988)\citenamefont{Chernikov,
  Sagdeev, and Zaslavsky}}]{kho_zaslavsky}
\bibinfo{author}{\bibfnamefont{A.~A.} \bibnamefont{Chernikov}},
  \bibinfo{author}{\bibfnamefont{R.~Z.} \bibnamefont{Sagdeev}},
  \bibnamefont{and} \bibinfo{author}{\bibfnamefont{G.~M.}
  \bibnamefont{Zaslavsky}}, \bibinfo{journal}{Physica D}
  \textbf{\bibinfo{volume}{33}}, \bibinfo{pages}{65} (\bibinfo{year}{1988}).

\bibitem[{\citenamefont{Vasiliev et~al.}(1989)\citenamefont{Vasiliev,
  Zaslavsky, Natenzon, Neishtadt, Petrovichev, Sagdeev, and
  Chernikov}}]{kho_diss}
\bibinfo{author}{\bibfnamefont{A.~A.} \bibnamefont{Vasiliev}},
  \bibinfo{author}{\bibfnamefont{G.~M.} \bibnamefont{Zaslavsky}},
  \bibinfo{author}{\bibfnamefont{M.~Y.} \bibnamefont{Natenzon}},
  \bibinfo{author}{\bibfnamefont{A.~I.} \bibnamefont{Neishtadt}},
  \bibinfo{author}{\bibfnamefont{B.~A.} \bibnamefont{Petrovichev}},
  \bibinfo{author}{\bibfnamefont{R.~Z.} \bibnamefont{Sagdeev}},
  \bibnamefont{and} \bibinfo{author}{\bibfnamefont{A.~A.}
  \bibnamefont{Chernikov}}, \bibinfo{journal}{JETP}
  \textbf{\bibinfo{volume}{67}}, \bibinfo{pages}{2053} (\bibinfo{year}{1989}).

\bibitem[{\citenamefont{Lasota and Mackey}(1994)}]{lasota}
\bibinfo{author}{\bibfnamefont{A.}~\bibnamefont{Lasota}} \bibnamefont{and}
  \bibinfo{author}{\bibfnamefont{M.~C.} \bibnamefont{Mackey}},
  \emph{\bibinfo{title}{Chaos, Fractals and Noise}}
  (\bibinfo{publisher}{Springer-Verlag}, \bibinfo{address}{New York},
  \bibinfo{year}{1994}).

\bibitem[{\citenamefont{Fox}(1995)}]{fox}
\bibinfo{author}{\bibfnamefont{R.~F.} \bibnamefont{Fox}},
  \bibinfo{journal}{Chaos} \textbf{\bibinfo{volume}{5}}, \bibinfo{pages}{619}
  (\bibinfo{year}{1995}).

\bibitem[{\citenamefont{Lichtenberg and Lieberman}(1994)}]{lieberman}
\bibinfo{author}{\bibfnamefont{A.~J.} \bibnamefont{Lichtenberg}}
  \bibnamefont{and} \bibinfo{author}{\bibfnamefont{M.~A.}
  \bibnamefont{Lieberman}}, \emph{\bibinfo{title}{Regular and Chaotic
  Dynamics}} (\bibinfo{publisher}{Springer-Verlag}, \bibinfo{address}{New
  York}, \bibinfo{year}{1994}).

\bibitem[{\citenamefont{Alan~Wolf and Vastano}(1985)}]{lyap}
\bibinfo{author}{\bibfnamefont{H.~L.~S.} \bibnamefont{Alan~Wolf},
  \bibfnamefont{J.~B.~Swift}} \bibnamefont{and}
  \bibinfo{author}{\bibfnamefont{J.~A.} \bibnamefont{Vastano}},
  \bibinfo{journal}{Physica D} \textbf{\bibinfo{volume}{16}},
  \bibinfo{pages}{285} (\bibinfo{year}{1985}).

\bibitem[{\citenamefont{Wineland et~al.}(1998)\citenamefont{Wineland, Monroe,
  Itano, Leibfried, King, and Meekhof}}]{wineland}
\bibinfo{author}{\bibfnamefont{D.~J.} \bibnamefont{Wineland}},
  \bibinfo{author}{\bibfnamefont{C.}~\bibnamefont{Monroe}},
  \bibinfo{author}{\bibfnamefont{W.~M.} \bibnamefont{Itano}},
  \bibinfo{author}{\bibfnamefont{D.}~\bibnamefont{Leibfried}},
  \bibinfo{author}{\bibfnamefont{B.~E.} \bibnamefont{King}}, \bibnamefont{and}
  \bibinfo{author}{\bibfnamefont{D.~M.} \bibnamefont{Meekhof}},
  \bibinfo{journal}{J. Res. Natl. Inst. Stand. Technol.}
  \textbf{\bibinfo{volume}{103}}, \bibinfo{pages}{29} (\bibinfo{year}{1998}).

\bibitem[{\citenamefont{Lindblad}(1976)}]{lindblad1}
\bibinfo{author}{\bibfnamefont{G.}~\bibnamefont{Lindblad}},
  \bibinfo{journal}{Math. Phys.} \textbf{\bibinfo{volume}{48}},
  \bibinfo{pages}{119} (\bibinfo{year}{1976}).

\bibitem[{\citenamefont{Alicki and Lendi}(1987)}]{lindblad2}
\bibinfo{author}{\bibfnamefont{R.}~\bibnamefont{Alicki}} \bibnamefont{and}
  \bibinfo{author}{\bibfnamefont{K.}~\bibnamefont{Lendi}},
  \emph{\bibinfo{title}{Quantum Dynamical Semigroups and Applications}}, vol.
  \bibinfo{volume}{286} of \emph{\bibinfo{series}{Lecture Notes in Physics}}
  (\bibinfo{publisher}{Springer-Verlag}, \bibinfo{address}{Berlin},
  \bibinfo{year}{1987}).

\bibitem[{\citenamefont{Carvalho et~al.}(2001)\citenamefont{Carvalho, Milman,
  de~Matos~Filho, and Davidovich}}]{arrc}
\bibinfo{author}{\bibfnamefont{A.~R.~R.} \bibnamefont{Carvalho}},
  \bibinfo{author}{\bibfnamefont{P.}~\bibnamefont{Milman}},
  \bibinfo{author}{\bibfnamefont{R.~L.} \bibnamefont{de~Matos~Filho}},
  \bibnamefont{and}
  \bibinfo{author}{\bibfnamefont{L.}~\bibnamefont{Davidovich}},
  \bibinfo{journal}{Phys. Rev. Lett.} \textbf{\bibinfo{volume}{86}},
  \bibinfo{pages}{4988} (\bibinfo{year}{2001}).

\bibitem[{\citenamefont{James}(1998)}]{james}
\bibinfo{author}{\bibfnamefont{D.~F.~V.} \bibnamefont{James}},
  \bibinfo{journal}{Phys. Rev. Lett.} \textbf{\bibinfo{volume}{81}},
  \bibinfo{pages}{317} (\bibinfo{year}{1998}).

\bibitem[{\citenamefont{Beck and Schl{\"o}gl}(1993)}]{beck}
\bibinfo{author}{\bibfnamefont{C.}~\bibnamefont{Beck}} \bibnamefont{and}
  \bibinfo{author}{\bibfnamefont{F.}~\bibnamefont{Schl{\"o}gl}},
  \emph{\bibinfo{title}{Thermodynamics of chaotic systems: an introduction}}
  (\bibinfo{publisher}{Cambridge University Press}, \bibinfo{address}{London},
  \bibinfo{year}{1993}).

\bibitem[{\citenamefont{Leibfried et~al.}(1996)\citenamefont{Leibfried,
  Meekhof, King, Monroe, Itano, and Wineland}}]{wigions}
\bibinfo{author}{\bibfnamefont{D.}~\bibnamefont{Leibfried}},
  \bibinfo{author}{\bibfnamefont{D.~M.} \bibnamefont{Meekhof}},
  \bibinfo{author}{\bibfnamefont{B.~E.} \bibnamefont{King}},
  \bibinfo{author}{\bibfnamefont{C.}~\bibnamefont{Monroe}},
  \bibinfo{author}{\bibfnamefont{W.~M.} \bibnamefont{Itano}}, \bibnamefont{and}
  \bibinfo{author}{\bibfnamefont{D.~J.} \bibnamefont{Wineland}},
  \bibinfo{journal}{Phys. Rev. Lett.} \textbf{\bibinfo{volume}{77}},
  \bibinfo{pages}{4281} (\bibinfo{year}{1996}).

\bibitem[{\citenamefont{Barnett and Radmore}(1997)}]{mtqo}
\bibinfo{author}{\bibfnamefont{S.~M.} \bibnamefont{Barnett}} \bibnamefont{and}
  \bibinfo{author}{\bibfnamefont{P.~M.} \bibnamefont{Radmore}},
  \emph{\bibinfo{title}{Methods in Theoretical Quantum Optics}}
  (\bibinfo{publisher}{Oxford university Press}, \bibinfo{address}{Oxford},
  \bibinfo{year}{1997}).

\bibitem[{\citenamefont{Zurek}(2003)}]{Zurek2}
\bibinfo{author}{\bibfnamefont{W.~H.} \bibnamefont{Zurek}},
  \bibinfo{journal}{Rev. Mod. Phys.} \textbf{\bibinfo{volume}{75}},
  \bibinfo{pages}{715} (\bibinfo{year}{2003}).

\bibitem[{\citenamefont{Gu}(1990)}]{lyapdist1}
\bibinfo{author}{\bibfnamefont{Y.}~\bibnamefont{Gu}}, \bibinfo{journal}{Phys.
  Lett. A} \textbf{\bibinfo{volume}{149}}, \bibinfo{pages}{95}
  (\bibinfo{year}{1990}).

\bibitem[{\citenamefont{Pattanayak and Brumer}(1997)}]{lyapdist2}
\bibinfo{author}{\bibfnamefont{A.~K.} \bibnamefont{Pattanayak}}
  \bibnamefont{and} \bibinfo{author}{\bibfnamefont{P.}~\bibnamefont{Brumer}},
  \bibinfo{journal}{Phys. Rev. E} \textbf{\bibinfo{volume}{56}},
  \bibinfo{pages}{5174} (\bibinfo{year}{1997}).

\end{thebibliography}
\end{document}